\pdfoutput=1
\documentclass[twocolumn,twocolappendix]{aastex631}
\usepackage{afterpage}
\usepackage{multirow}
\usepackage{listings}
\usepackage[tbtags]{amsmath}
\usepackage{hyperref}
\usepackage{breqn}
\usepackage[normalem]{ulem} 
\usepackage{csquotes}
\usepackage{courier}
\MakeOuterQuote{"}

\def\arcsec{\hbox{$^{\prime\prime}$}}

\definecolor{mypink1}{rgb}{0.858, 0.188, 0.478}
\definecolor{mypink2}{rgb}{0.99, 0.45, 0.75}
\definecolor{mygreen}{rgb}{0.0, 0.5, 0.1}

\defcitealias{arnaud2010}{A10}
\defcitealias{comis2011}{C11}
\defcitealias{marrone2012}{M12}
\defcitealias{planelles2017}{P17}
\defcitealias{vikhlinin2006b}{V06}

\begin{document}

\title{Inferences from surface brightness fluctuations of Zwicky 3146 via the Sunyaev-Zel'dovich effect and X-ray observations}
\author[0000-0001-5725-0359]{Charles E. Romero}
\altaffiliation{E-mail: \href{mailto:charles.romero@cfa.harvard.edu}{charles.romero@cfa.harvard.edu} }
\affiliation{Center for Astrophysics $\vert$ Harvard \& Smithsonian, 60 Garden Street, Cambridge, MA 02138, USA}

\author[0000-0003-2754-9258]{Massimo Gaspari}
\affiliation{Department of Astrophysical Sciences, Princeton University, 4 Ivy Lane, Princeton, NJ 08544-1001, USA}

\author{Gerrit Schellenberger}
\affiliation{Center for Astrophysics $\vert$ Harvard \& Smithsonian, 60 Garden Street, Cambridge, MA 02138, USA}

\author{Tanay Bhandarkar}
\affiliation{Department of Physics and Astronomy, University of Pennsylvania, 209 South 33rd Street, Philadelphia, PA, 19104, USA}

\author[0000-0002-3169-9761]{Mark Devlin}
\affiliation{Department of Physics and Astronomy, University of Pennsylvania, 209 South 33rd Street, Philadelphia, PA, 19104, USA}

\author[0000-0002-1940-4289]{Simon R.\ Dicker} 
      \affiliation{Department of Physics and Astronomy, University of Pennsylvania, 209 South 33rd Street, Philadelphia, PA, 19104, USA}

\author{William Forman}
\affiliation{Center for Astrophysics $\vert$ Harvard \& Smithsonian, 60 Garden Street, Cambridge, MA 02138, USA}

\author{Rishi Khatri}
\affiliation{Tata Institute of Fundamental Research, Homi Bhabha Road, Mumbai 400005, India}

\author{Ralph Kraft}
\affiliation{Center for Astrophysics $\vert$ Harvard \& Smithsonian, 60 Garden Street, Cambridge, MA 02138, USA}

\author[0000-0003-3586-4485]{Luca Di Mascolo}
\affiliation{Department of Physics, University of Trieste, via Tiepolo 11, 34131 Trieste, Italy}
\affiliation{INAF - Osservatorio Astronomico di Trieste, via Tiepolo 11, 34131 Trieste, Italy}
\affiliation{IFPU - Institute for Fundamental Physics of the Universe, Via Beirut 2, 34014 Trieste, Italy}

\author[0000-0002-8472-836X]{Brian S.\ Mason}
\affiliation{National Radio Astronomy Observatory, 520 Edgemont Rd., Charlottesville VA 22903, USA}

\author[0000-0001-9793-5416]{Emily Moravec}
\affiliation{Green Bank Observatory, 155 Observatory Road, Green Bank, WV 24944, USA}

\author[0000-0003-3816-5372]{Tony Mroczkowski}
\affiliation{ESO - European Southern Observatory, Karl-Schwarzschild-Str.\ 2, D-85748 Garching b.\ M\"unchen, Germany}

\author{Paul Nulsen}
\affiliation{Center for Astrophysics $\vert$ Harvard \& Smithsonian, 60 Garden Street, Cambridge, MA 02138, USA}
\affiliation{ICRAR, University of Western Australia, 35 Stirling Hwy, Crawley, WA 6009, Australia}

\author[0000-0003-1842-8104]{John Orlowski-Scherer}
\affiliation{Department of Physics and Astronomy, University of Pennsylvania, 209 South 33rd Street, Philadelphia, PA, 19104, USA}

\author{Karen Perez Sarmiento}
\affiliation{Department of Physics and Astronomy, University of Pennsylvania, 209 South 33rd Street, Philadelphia, PA, 19104, USA}

\author[0000-0003-0167-0981]{Craig Sarazin} 
\affiliation{Department of Astronomy, University of Virginia,  530 McCormick Road, Charlottesville, VA 22901, USA}

\author[0000-0001-6903-5074]{Jonathan Sievers}
\affiliation{Department of Physics, McGill University, 3600 University Street Montreal, QC H3A 2T8, Canada}

\author{Yuanyuan Su}
\affiliation{Department of Physics and Astronomy, University of Kentucky, 505 Rose Street, Lexington, KY 40506, USA}


\shorttitle{Fluctuations in Zwicky 3146}
\shortauthors{Romero et al.}
\lstset{basicstyle=\ttfamily,language=bash}

\begin{abstract}
The galaxy cluster Zwicky 3146 is a sloshing cool core cluster at $z{=}0.291$ that in SZ imaging does not appear to exhibit significant pressure substructure in the intracluster medium (ICM). 
We perform a surface brightness fluctuation analysis via Fourier amplitude spectra on SZ (MUSTANG-2) and X-ray (\textit{XMM-Newton}) images of this cluster. These surface brightness fluctuations can be deprojected to infer pressure and density fluctuations from the SZ and X-ray data, respectively. In the central region (Ring 1, $r < 100\arcsec = 440$ kpc, in our analysis) we find fluctuation spectra that suggest injection scales around 200 kpc ($\sim 140$ kpc from pressure fluctuations and $\sim 250$ kpc from density fluctuations). 
When comparing the pressure and density fluctuations in the central region, we observe a change in the effective thermodynamic state from large to small scales, from isobaric (likely due to the slow sloshing) to adiabatic (due to more vigorous motions). By leveraging scalings from hydrodynamical simulations, we find an average 3D Mach number $\approx0.5$.
We further compare our results to other studies of Zwicky 3146 and, more broadly, to other studies of fluctuations in other clusters.

\end{abstract}


\section{Introduction}
\label{sec:intro}

The dominant baryonic component of galaxy clusters is the hot intracluster medium (ICM) which can be observed via X-rays and in the millimeter band via the Sunyaev-Zel'dovich (SZ) effect \citep{sunyaev1970,sunyaev1972}. The observed radiative signatures at the two wavelengths regimes both depend on thermodynamic properties integrated along the line of sight (the gas is optically thin in both regimes), with X-ray surface brightness being roughly proportional to square of gas density integrated along the line of sight and the millimeter surface brightness being proportional to electron pressure along the line of sight. Temperatures can then be inferred from X-ray spectra or by combining pressure constraints from SZ data with density constraints from X-ray data \citep[e.g.][]{romero2017, bourdin2017}. 

Cluster masses can be estimated assuming hydrostatic equilibrium from radial profiles of gas density and profiles of either gas temperature or pressure. The mass inferred under the assumption of hydrostatic equilibrium is expected to fall below the true mass of the cluster by 10-30\% \citep[e.g.][]{hurier2018}. This offset from the true mass is termed "hydrostatic bias" and is expected to be primarily due to non-thermal pressure support, in particular turbulent motions driven by mergers and feedback tied to active galactic nuclei (AGN; \citealt{gaspari20}, for a review).

The extent to which the non-thermal pressure is dominated by velocity fluctuations of the gas can be revealed through Doppler broadening of emission lines observed by upcoming X-ray missions with high spectral resolution such as XRISM \citep{xrism2020} and \textit{Athena} \citep{athena2013,roncarelli2018}. 
Fluctuations in thermodynamic quantities may reveal the nature of hydrostatic bias. In particular, pressure fluctuations ($\delta P / P$) can quantify the relative non-thermal pressure of the thermal gas\footnote{It is often expected that (quasi) turbulent motions dominate the non-thermal pressure, though cosmic rays and magnetic fields may also contribute to the non-thermal pressure.}.

\begin{figure}[!t]
    \centering
    \includegraphics[width=0.45\textwidth]{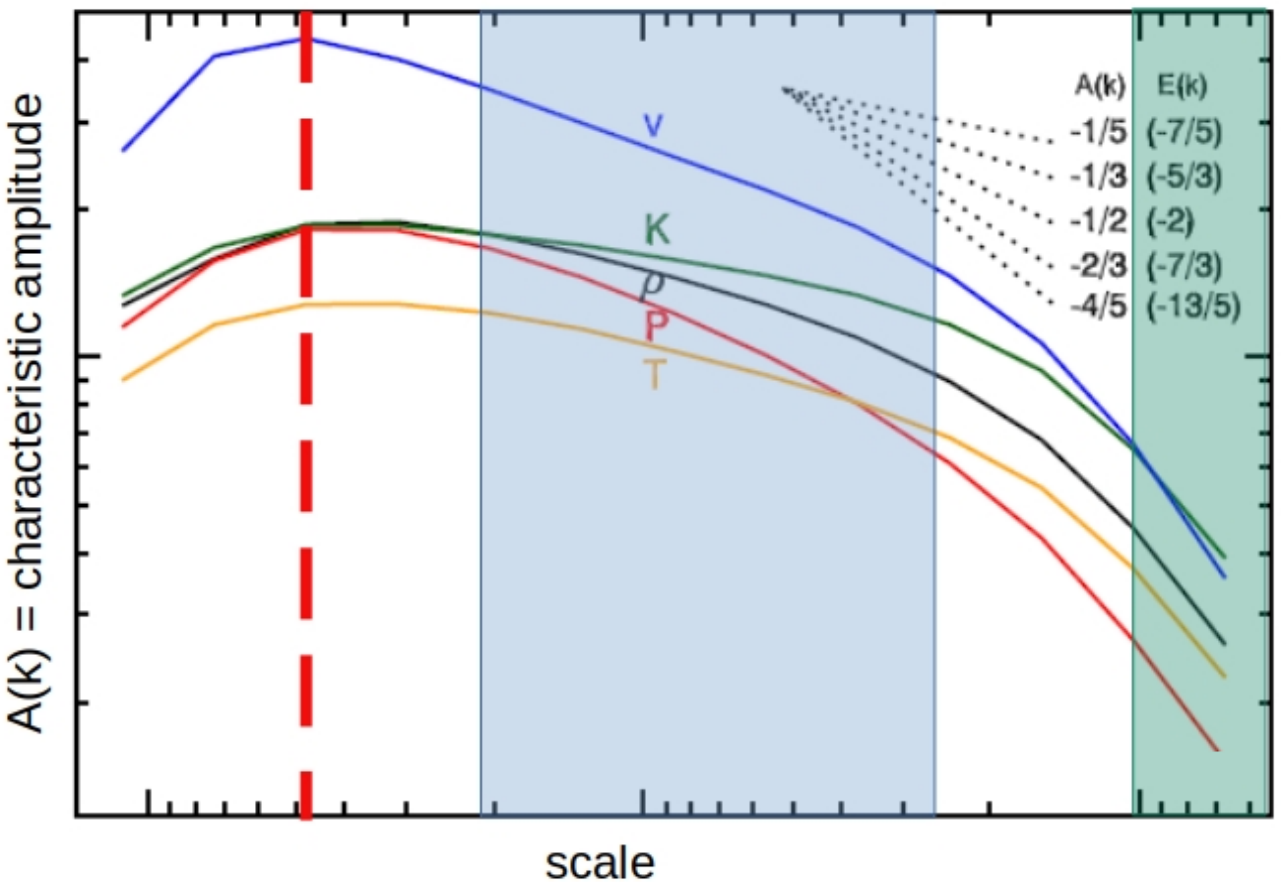}
    \caption{Figure adapted from \citet{Gaspari2014_PS} showing typical ICM amplitude spectra for the thermodynamic relative fluctuations: density ($\delta \rho / \rho$), temperature ($\delta T / T$), entropy ($\delta K / K$), pressure ($\delta P / P$), and velocity $\delta v / c_s$. Smaller scales (distances) are towards the right of the plot; values are suppressed to allow for generalization, i.e. for an arbitrary injection scale, we expect the same shape (roughly) for the spectra, with the peak of the spectra at said injection scale. The red dashed line indicates the injection scale; the shaded blue region indicates the scales over which the fluctuations "cascade" towards smaller scales, and the shaded green region is where the fluctuations are finally dissipated. The dotted black lines help guide the eye as to the (logarithmic) slope of the various spectra, which again should not be treated as an exact expectation; the slopes will vary depending on the actual conditions of the ICM in a given cluster.}
    \label{fig:toyspectrum}
\end{figure}

To quantify fluctuations as a function of scale, we use amplitude spectra leveraging the Fourier domain \citep[e.g.][]{Churazov2012,Gaspari2013_PS,Gaspari2014_PS}. As in previous studies, the amplitude spectrum is defined as
\begin{equation}
    A(k) \equiv \sqrt{P(k) 4 \pi k^3},
\end{equation}
where $k = \sqrt{k_x^2 + k_y^2 + k_z^2}$ and $P(k)$ is the power spectrum. Figure~\ref{fig:toyspectrum} has been adapted from \citet{Gaspari2014_PS} to highlight key features/regions of interest in the amplitude spectra of thermodynamic fluctuations or velocity fluctuations ($\delta v / c_s$, where $c_s$ is the sound speed) when considering a single dominant injection mechanism. In particular, Figure~\ref{fig:toyspectrum} illustrates three important length scales (or range of scales): an injection scale, $l_{\text{inj}}$ (e.g.~for mergers, expected to be several hundreds of kiloparsecs), intermediate scales ($\sim$10-100 kpc) at which the fluctuations are "cascading" towards smaller scales, and small scales at which the fluctuations are gradually dissipated, e.g.~via Coulomb collisions or Alfv\'{e}n/whistler waves \citep[e.g.][]{drake2021,cho2022}. The values in Figure~\ref{fig:toyspectrum} are suppressed to allow for generalization, i.e. the injection scale used in the particular simulation(s) may not match those in a particular cluster, e.g. Zwicky 3146, but we still expect the same general shape of the amplitude spectra (or the summation of such spectra if there are multiple injection mechanisms). The amplitude of the relevant fluctuations is generally taken as the maximum of the amplitude spectrum, $A_{\text{3D}}(k_{\text{peak}})$. The scales at which the damping occurs is generally expected to be smaller than can be (spatially) resolved for most galaxy clusters.  

Most of the previous studies focused on retrieving the amplitude spectrum of a galaxy cluster using solely X-ray observations \citep[e.g.][]{schuecker2003,Churazov2012,sanders2012,gaspari13,Zhuravleva2014b,arevalo2016}. 
Similar studies have also targeted the amplitude/variance of fluctuations \citep[e.g.][]{hofmann2016,eckert2017}.
However, pure X-ray observations are often limited to less than a decade in spatial scale, and mostly targeting density fluctuations.
To overcome such limitations, a multiwavelength approach is required. As a first exploratory study, \citet{khatri2016} showed that SZ images (via \textit{Planck}) are a key complementary tool to X-ray datasets, in particular expanding our knowledge of relative ICM fluctuations over the large scales (low Fourier $k$ modes) and the pressure variable. 
Here, we continue such a multiwavelength approach by leveraging the capabilites of MUSTANG-2.

In this paper we present a study of surface brightness fluctuations of SZ and X-ray maps of Zwicky 3146, also referred to as ZwCl 1021.0+0426, and associated amplitude spectra covering a decade in scales. Zwicky 3146 \citep[$z=0.291$,][]{allen1992} is a massive, relaxed, sloshing cluster with a cool core \citep{forman2002}. The relaxed and regular nature of Zwicky 3146 give us the expectation that we will not find large pressure fluctuations. This work is a follow-up work to the study of Zwicky 3146 presented in \citet{romero2020} (wherein Zwicky 3146 is also described in more detail). In particular, \citet{romero2020} estimated the mass of Zwicky 3146 from pressure profiles determined from high-resolution SZ data and varying assumptions, including hydrostatic equilibrium when combined with electron density profiles determined from \textit{XMM-Newton} data. Masses from \citet{romero2020} and references therein \citep[e.g.][]{klein2019,hilton2018,martino2014} are in agreement with $M_{500} = 8 \times 10^{14} M_{\odot}$, which corresponds to $R_{500} = 5$ arcminutes (1.3 Mpc).


The layout of this paper is as follows. Section~\ref{sec:data} describes the data used and fitted surface brightness models. To perform our fluctuation analysis, detailed in Section~\ref{sec:PSMethod}, we calculate power spectra on fractional residual maps; that is, residual maps divided by their respective surface brightness models. We present the 2D and (deprojected) 3D amplitude spectra
in Section~\ref{sec:results} and discuss them in the context of what is known about Zwicky 3146 in Section~\ref{sec:discussion}. We offer conclusions in Section~\ref{sec:conclusions}.

Throughout this paper, we adopt a concordance cosmology: $H_0 = 70$~km~s$^{-1}$~Mpc$^{-1}$, $\Omega_M = 0.3$, $\Omega_{\Lambda} = 0.7$. We define $h_{70} \equiv H_0$~(70 km~s$^{-1}$~Mpc$^{-1}$)$^{-1}$ and $h(z) \equiv H(z) H_0^{-1}$. At the redshift of Zwicky 3146 ($z=0.291$), one arcsecond corresponds to 4.36~kpc.


\section{Data products}
\label{sec:data}
   We make use of MUSTANG-2 data presented in \citet{romero2020} and archival \textit{XMM-Newton} EPIC data.
The two data sets are highly complementary. MUSTANG-2 has a resolution (full-width half maximum, FWHM) of $\sim10^{\prime\prime}$. The PSF of each of \textit{XMM-Newton}'s detectors, MOS1, MOS2, and pn, depends on the energy, and off-axis distance; for a point of rough comparison, we may consider that the detectors have an effective resolution of $\sim5^{\prime\prime}$, albeit with broad wings.

\subsection{MUSTANG-2 data products}
\label{sec:Mustang_data}

    MUSTANG-2 is a 215-detector array on the 100-m Robert C. Byrd Green Bank Telescope (GBT) and achieves $10^{\prime\prime}$ resolution (FWHM) with an instantaneous field of view (FOV) of $4^{\prime}.2$.
    Observing at 90 GHz, it is sensitive to the SZ effect, which is often parameterized in terms of Compton $y$:
    \begin{equation}
        y = \frac{\sigma_{\text{T}}}{m_{\text{e}} c^2} \int P_e(\theta,z) dz,
        \label{eqn:comptony}
    \end{equation}
    where $\sigma_T$ is the Thomson cross section, $m_e$ is the electron mass, $c$ the speed of light, $P_e$ the electron pressure, and $z$ is the axis along the line of sight.

     \begin{figure}[h]
        \begin{center}
        \includegraphics[width=0.49\textwidth]{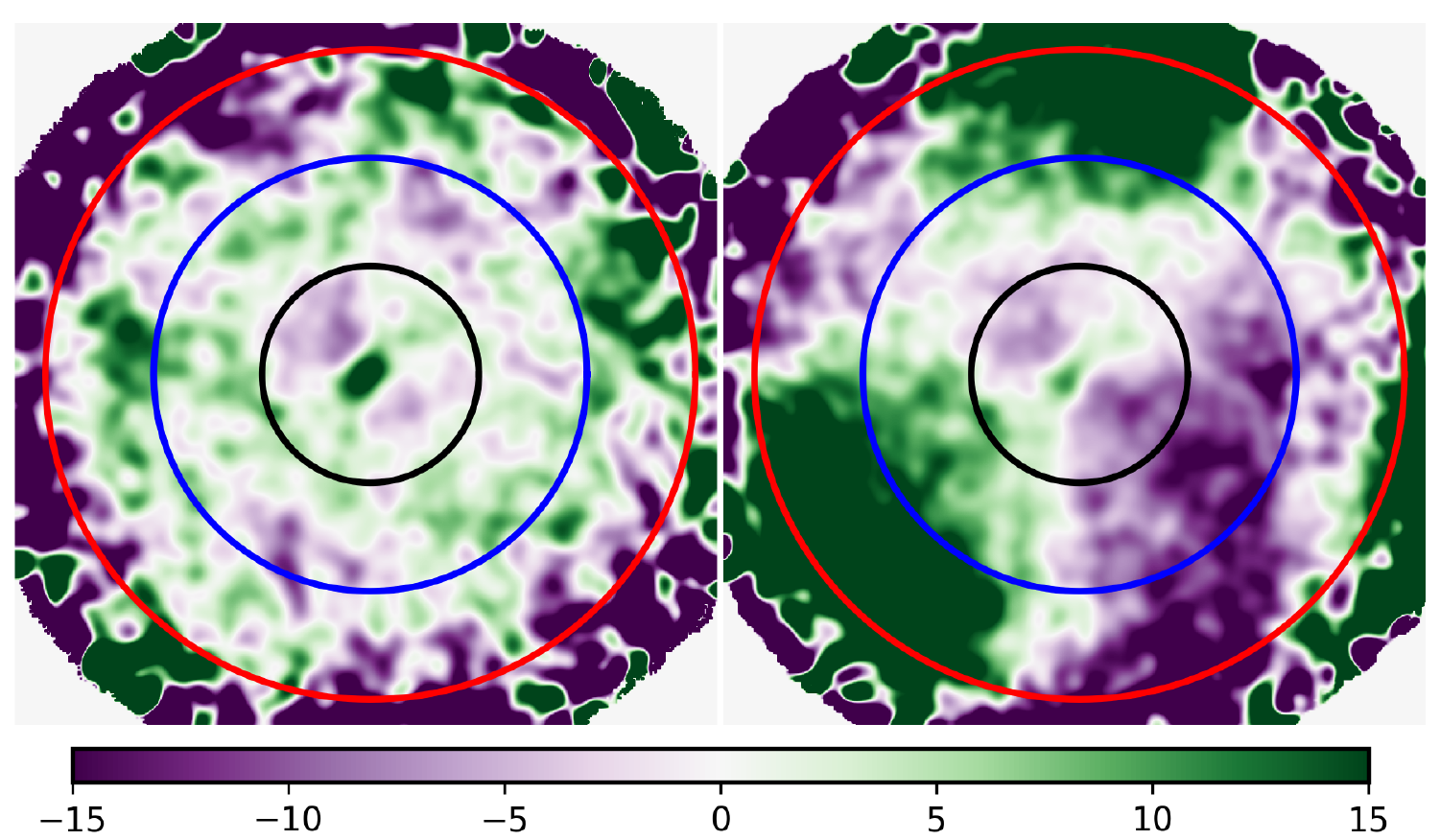}
        \end{center}
        \caption{Maps derived from the MUSTANG-2 observations: the residual Minkasi map (right) shows large scale noise, while the residual MIDAS map (left) has this filtered out. Given the angular scales of interest, the MIDAS map is preferable. The rings are as in Figure~\ref{fig:fract_resids}. The color scale is shown in units of $y \times 10^{6}$; $y$ is defined in Equation~\ref{eqn:comptony}.
        }
        \label{fig:deltay_maps}
    \end{figure}
    
    The observations used here are the same as in \citet{romero2020}, as is the general data reduction. We employ both data reduction pipelines, MIDAS and Minkasi, in this work. Briefly, MIDAS follows a more traditional approach in its data processing (i.e. similar to the processing of many predecessor multi-pixel bolometric ground-based measurements); this processing typically restricts scales recovered (often characterized as a transfer function)\footnote{The transfer function as used in \citet{romero2020} is quantified as the transmission of the Fourier transform of an input map.} to less than the instrument's instantaneous FOV (see Figure~\ref{fig:deltay_maps}.) Meanwhile, Minkasi fits the data in the time domain and does not suffer the same loss of scales as MIDAS; see \citet{romero2020} for a detailed comparison of the transfer functions of the two processing methods.

     \begin{figure}[!h]
        \begin{center}
        \includegraphics[width=0.45\textwidth]{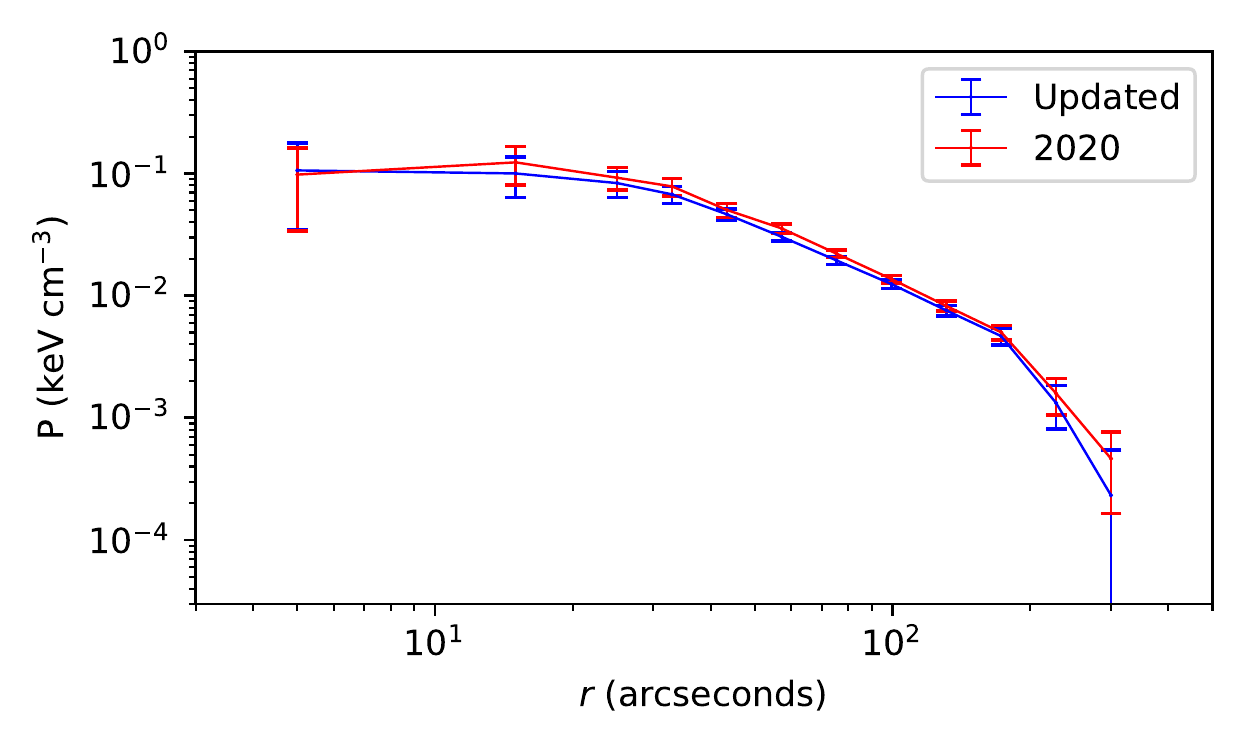}       \end{center}
        \caption{Our updated profile is consistent with our previously published profile; we do see the outermost bin has a lower pressure than previously \citep{romero2020}.
        }
        \label{fig:radial_profiles}
    \end{figure}
   
    In this work, we update our pressure profile model from \citet{romero2020} with an additional procedure used in \citet{dicker2020,orlowski2022} which attempts to further remove atmospheric contributions to our maps by fitting and subtracting a second order polynomial with respect to elevation offset from the scan center. Figure ~\ref{fig:radial_profiles} compares the current to the former pressure profile; the two are fully consistent with each other.



    As reported in \citet{romero2020}, the two pressure profile models (fit via MIDAS and Minkasi) are consistent, except beyond MUSTANG-2's radial (instantaneous) FOV where our transfer function is poorly constrained. However, when we subtract the Minkasi model via the MIDAS pipeline (rather than using a transfer function), we see that the residual map is consistent with noise at the radii where the pressure profiles (MIDAS vs Minkasi) differ.


\subsection{\textit{XMM} data products and models}
\label{sec:XMM_data}
   There are four \textit{XMM-Newton} observations (Obs.IDs) of Zwicky 3146: 0108670401, 0108670101, 0605540301, and 0605540201. The first does not have usable EPIC data; we use the remaining three observations (of nominal durations 56, 65, and 123 ks; see also Table~\ref{tbl:xmm_obs}).

We use heasoft v6.28 and SAS 19.0 and the Extended Source Analysis Software (ESAS) data reduction package \citep{snowden2008} to produce event files and eventually  images for the three EPIC detectors: MOS1, MOS2, and pn. Our data reduction largely follows the ESAS cookbook\footnote{\url{https://heasarc.gsfc.nasa.gov/docs/xmm/esas/cookbook/xmm-esas.html}}, with the initial steps being \lstinline{emchain}, \lstinline{epchain}, and \lstinline{epchain withoutoftime=true} to extract calibrated events files. Soft proton flares are excised with the tasks \lstinline{mos-filter} and \lstinline{pn-filter}. A comparison of IN versus OUT count rates assesses the amount of residual contamination from soft protons \citep{deluca2004}. This comparison suggests that soft protons are not a concern for MOS detectors and that the pn detectors could suffer slight contamination.


\begin{table}[]
    \centering
    \begin{tabular}{c|c|c|c}
         Obs ID & 0108670101 & 0605540301 & 0605540201 \\
         \hline
         Date & 2000 Dec 05 & 2009 May 08 & 2009 Dec 13 \\
         Exposure (ks) & 56.5 & 64.9 & 122.8 \\
         \hline
         \multirow{3}{*}{Clean Exp (ks)} & MOS1: 51.2 & MOS1: 41.8 & MOS1: 101.2 \\
          & MOS2: 51.7 & MOS2: 40.6 & MOS2: 102.2 \\
          & pn: 43.3 & pn: 29.6 & pn: 73.8 \\
          \hline
         Mode &  FF & eFF & eFF \\
         PI & R. Mushotzky & J. Sanders & J. Sanders
    \end{tabular}
    \caption{Overview of imaging \textit{XMM-Newton} observations of Zwicky 3146. Modes FF and eFF are "full frame" and "extended full frame", respectively.}

    \label{tbl:xmm_obs}
\end{table}

An initial list of point sources is created with the task  \lstinline{cheese} on the \textit{XMM-Newton} dataset based on flux with [0.4-7.2] keV energy band and detection significance. A region file is generated, excluding a $30^{\prime\prime}$ radius about each point source.

\subsubsection{Image creation}
\label{sec:img}

We choose to extract images in the [0.4-1.25] keV and [1.25-5.0] keV bands. Images and vignetted exposures are extracted for each detector over the entire detector area whilst masking point sources (see Section~\ref{sec:ptsrc_exc} for point source identification) via the task \lstinline{mos-spectra} or \lstinline{pn-spectra}. Unvignetted exposures are also created with the task \lstinline{eexpmap withvignetting=no}. Wide band (i.e. [0.4-5.0] keV) images are formed by the simple addition of the counts (and background) images; these wide band images are used for consistency checks.

\subsubsection{Constrained background components}
\label{sec:cbc}

The relevant particle backgrounds are calculated for the desired energy band via the tasks \lstinline{mos_back} and \lstinline{pn_back}.  For the pn detector, we extract a separate spectrum (via \lstinline{pn-spectra}) over the cluster region, which we take to be a radius of 5 arcminutes about the cluster center. While we treat the residual soft proton spectrum as a single power law, we must fit several other components to the spectrum: a thermal plasma component (\lstinline{apec}) for each of the local (Solar) hot bubble, Galactic emission, and the ICM in Zwicky 3146. In addition to this, we also consider Gaussian components for fluorescent lines. A soft proton background is then made with the task \lstinline{proton} and added to the particle background with the task \lstinline{farith}. For the pn detector, we also consider the out-of-time (OOT) contribution. Depending on the full frame mode, we multiply our resultant pn image with randomized columns by 0.063 or 0.023 for full frame and extended frame modes, respectively, to have an OOT component which we incorporate into the pn background. These background images will be subtracted from the respective images when extracting profiles.

\subsubsection{Point Source exclusion}
\label{sec:ptsrc_exc}

In addition to the list generated from \lstinline{cheese}, we make use of \textit{Chandra} archival data of Zwicky 3146 and run \lstinline{wavdetect} on its calibrated event files. Finally, we perform a manual inspection to identify any remaining point sources.

\subsubsection{Profile fitting}
\label{sec:xray_profile}

We use the Python package \lstinline{pyproffit} \citep{eckert2017} to extract profiles of our images.  Profiles are fit via \lstinline{emcee} \citep{foreman2013} separately for each detector, each energy band, and each observation.  We fit profiles to our low energy([0.4-1.25] keV) and high energy ([1.25-5.0] keV) images; As these profiles are fit per detector and per ObsID, we have 18 profiles in total (with another 9 from the wide energy band ([0.4-5.0] keV images that are only used for consistency checks).

    \begin{figure}[!h]
        \begin{center}
        \includegraphics[width=0.49\textwidth]{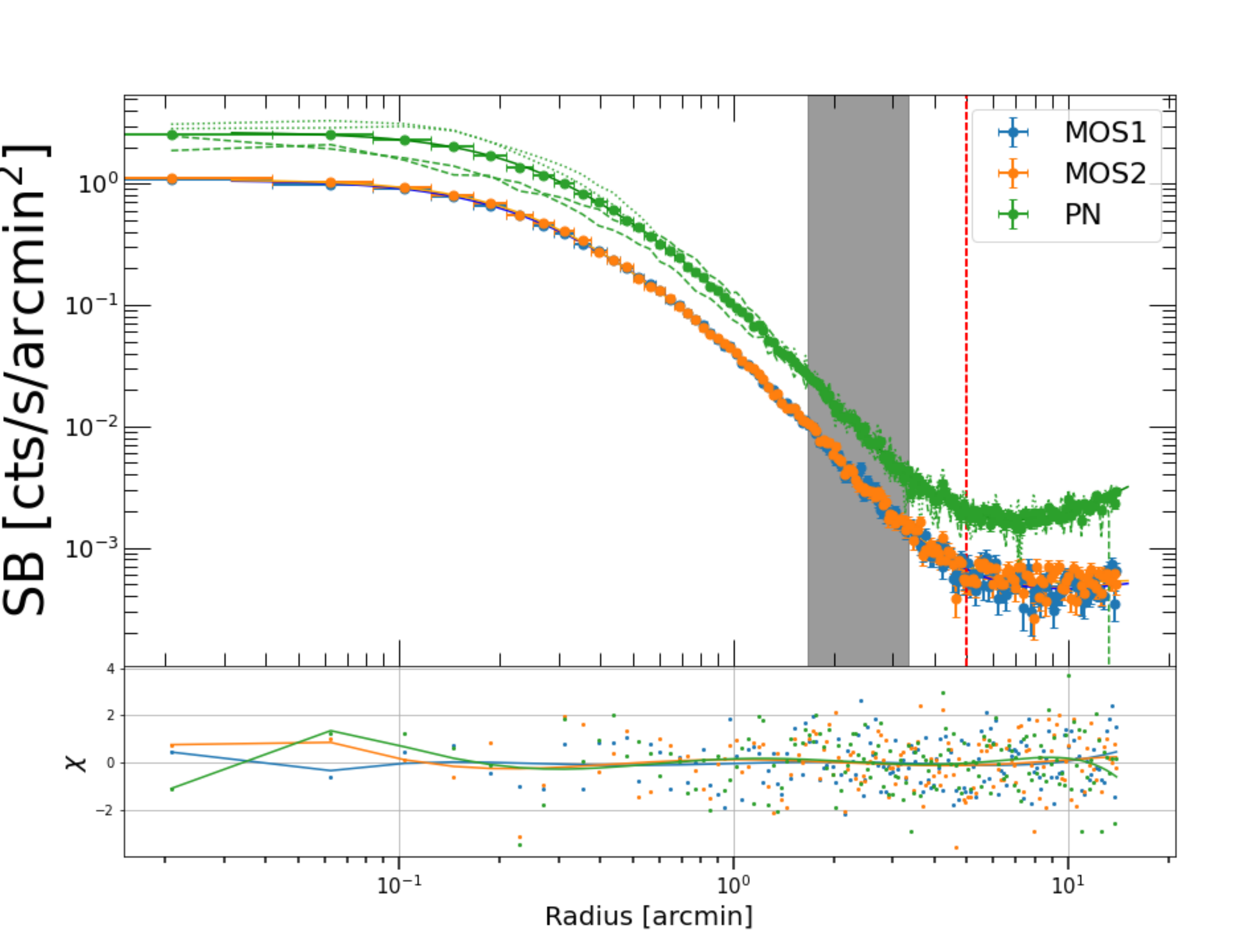}
        \end{center}
        \caption{Profile fits of circular double $\beta$-models to each detector array in our "high energy" (1250-5000 eV) band for observation ID 0605540201. The grey vertical band is between $100^{\prime\prime}$ and $200^{\prime\prime}$, i.e. the region used for Ring 2. The vertical red line is at $300^{\prime\prime}$ ($\sim R_{500}$ and the outer edge of Ring 3) The dotted and dashed green curves show the PN profile broken into quadrants (along cardinal directions); the dotted lines are the two western quadrants and the dashed lines are the eastern quadrants. The lines in the residual are a polynomial regression to indicate large-scale residuals.
        }
        \label{fig:xmm_sb_profiles}
    \end{figure}

Beyond masking the point sources, we also introduce a mask to exclude pixels of low exposure due to binning near chip gaps. We allow \lstinline{pyproffit} to fit for centroids in the central 5 arcminutes of each (masked) image independently. Within a single observation and energy band the centroids of each detector differ by $\lesssim2^{\prime\prime}$. Given the general agreement, for each observation and energy band, we adopt circular symmetry and the centroid as the average centroid of the maps from each EPIC camera detector when extracting profiles. To be sure, the centroids determined in this manner differ by $\sim 3\arcsec$ relative to the centroid used with MUSTANG-2 analysis. 

We find that a simple $\beta$-model does not sufficiently capture the surface brightness in the core of Zwicky 3146 and at large radii. We adopt the double $\beta$-model as implemented in \lstinline{pyproffit}, which has the form:
\begin{dmath}
    S(r) = S_0 [(1+(r/r_{c,1})^2)^{-3 \beta + 0.5} +  R(1+(r/r_{c,2})^2)^{-3 \beta + 0.5}] + B,
    \label{eqn:double_beta}
\end{dmath}
where $r$ is the radius, $r_{c,1}$ is the first "core" (scaling) radius, $r_{c,2}$ is the second "core" (scaling) radius, $R$ is a ratio between the two $\beta$-profile components, $S_0$ is the surface brightness normalization, and $B$ is the background. We modify the background component (taken to be uniform in \lstinline{pyproffit}) to be two components: one uniform and one the scaling of unvignetted-to-vignetted exposure maps. This latter component allows us to capture the contribution from fluorescent lines, predominantly the line from Aluminium, which is evident in the extracted profiles seen in Figure~\ref{fig:xmm_sb_profiles}.

To appropriately constrain these background components we find that we should fit (from $r=0$) out to at least 10 arcminutes, but beyond 10 arcminutes the values of the background components do not change much. We choose 11 arcminutes (more than $2 R_{500}$) as our fitting region. Across all three observation IDs, detectors, and energy bands, the profile residuals are quite small as in Figure~\ref{fig:xmm_sb_profiles}.

We find that the residuals of the double $\beta$-model are generally very small, with slightly larger residuals towards the core where known sloshing exists \citep[e.g.][]{forman2002}. We find that this is not a shortcoming of the double $\beta$-model per se but rather affirmation that the surface brightness of the cluster, while roughly circular at large radii, is not circular in the core \citep[cf axial ratios found in][]{romero2020}.

\section{Power spectra measurements}
\label{sec:PSMethod}
\begin{figure}[!h]
        \begin{center}
        \includegraphics[width=0.45\textwidth]{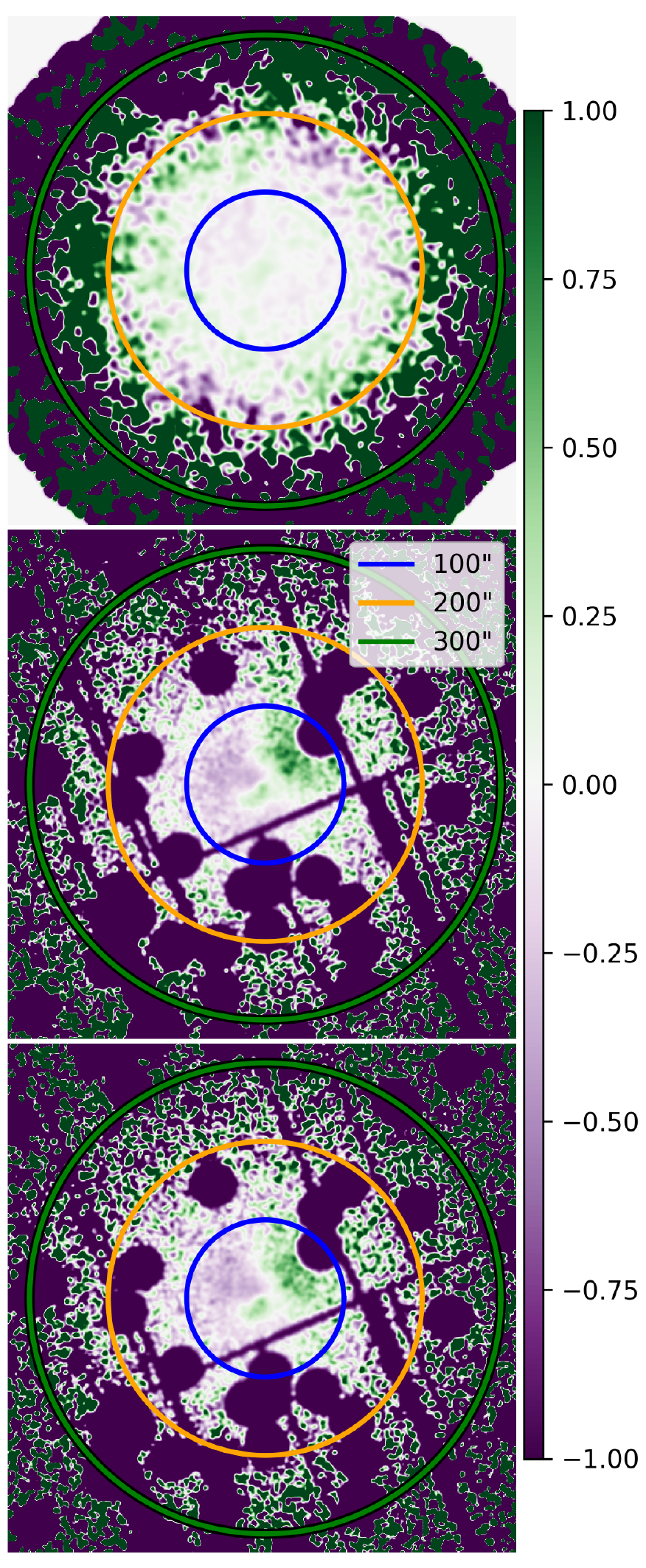}
        \end{center}
        \caption{Fractional residuals of our MUSTANG-2 data (upper) and \textit{XMM-Newton} data (only pn chip from observation 0605540201 shown) in the middle (400-1250 eV) and bottom (1250-5000 eV). The blue, orange, and green circles indicate $r = 100^{\prime\prime}$, $200^{\prime\prime}$, and $300^{\prime\prime}$, respectively. The purple lines and circles are masked chip gaps and point sources.
        }
        \label{fig:fract_resids}
    \end{figure}

    To quantify the fluctuations in surface brightness, we want to take the power spectra of residual images divided by the corresponding ICM surface brightness model as shown in Figure~\ref{fig:fract_resids}. We term these images "fractional residuals" and they are designated by either $\delta S/S$ for X-ray images or $\delta y/y$ for SZ images.  
In particular, Figure~\ref{fig:fract_resids} shows fractional residual maps for MUSTANG-2 and pn images from a single observation in the 400-1250 eV and 1250-5000 eV bands.
From these (2D) spectra of the images, we can deproject to spectra of underlying 3D thermodynamical quantities, namely pressure for SZ images and density for X-ray images (see Section~\ref{sec:3dpsmethod}).

Motivated in part by the data, as well as by the theoretical expectation for differing levels of fluctuations as a function of cluster-centric radii, we divide the cluster into three annuli: 
\begin{itemize}
    \item Ring 1: $r < 100^{\prime\prime} = 440$ kpc
    \item Ring 2: $100^{\prime\prime} < r < 200^{\prime\prime}$, and 
    \item Ring 3: $200^{\prime\prime} < r < 300^{\prime\prime} = R_{500}$.
\end{itemize}
We also note that the MUSTANG-2 map has a rapidly increasing RMS beyond $200^{\prime\prime}$ while the RMS is nearly uniform within $100^{\prime\prime}$.

We calculate the power spectra of the fractional residual images, $P_{\rm 2D}$ at five angular scales spaced logarithmically between $10^{\prime\prime}$ (the FWHM of MUSTANG-2) and $100^{\prime\prime}$ (the radial width of our annuli, i.e. rings). Corresponding amplitude spectra, $A_{2D}$ and $A_{3D}$ are given as
\begin{align}
    A_{2D}(k) &= [k^2 P_{\rm 2D}  * (2 \pi)]^{1/2} \label{eqn:as2d} \\
    A_{3D}(k) &= [k^3 P_{\rm 3D}  * (4 \pi)]^{1/2} \label{eqn:as3d}
\end{align}.

We use a modified $\Delta$-variance method \citep{arevalo2012} to calculate the power spectra of surface brightness fluctuations. In particular, this method allows us to recover power spectra of data with arbitrary gaps (masks) in (of) the data, which suits our needs well. We do, however, need to be cautious of the bias that can occur due to steep underlying spectra; this is especially true given that we will attempt to recover spectra up to scales close to the FWHM of MUSTANG-2 and \textit{XMM}. In particular, the convolution of a moderate slope with the PSF for either MUSTANG-2 or any of the EPIC cameras will lead to not only a steep slope, but a changing steep slope. The bias for this changing slope is derived in Appendix~\ref{sec:powerspectracircles}. While we report 
spectral values at $k = 0.1$ arcsec$^{-1}$ in later figures, this bias and associated uncertainty reduces the significance of the values at $k = 0.1$ arcsec$^{-1}$ such that none of them is statistically significant.



\subsection{Calculations on MUSTANG-2 data}
\label{sec:SZPS}

    As noted in Section~\ref{sec:Mustang_data}, our MUSTANG-2 residual map is created by subtracting the best fit model (from Minkasi) within the MIDAS pipeline. In all, 155 scans on source are used. Maps are produced for each scan, and the final residual image (see again Figure~\ref{fig:fract_resids}, top panel) is constructed as the (weighted) sum of these individual scan maps. 
    
\begin{figure}
        \begin{center}
        \includegraphics[width=0.45\textwidth]{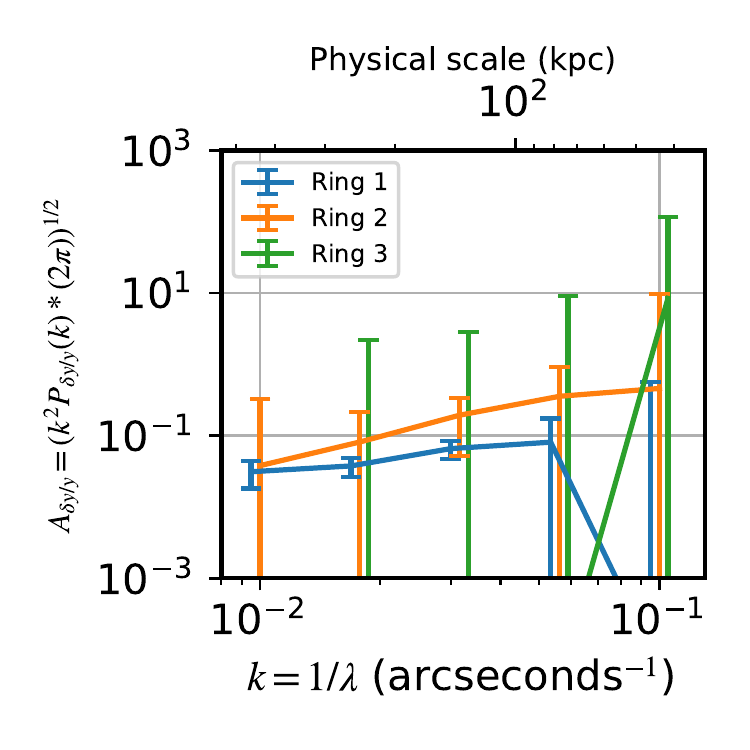}
        \end{center}
        \caption{The amplitude spectrum of the fractional residual ($\delta y / \bar{y}$) for each ring. Abscissa values are offset between rings for visual separation. Our best constraints are in Ring 1, while Ring 2 is already quite noisy.
        }
        \label{fig:dy_ms}
    \end{figure}

    In order to calculate power spectra due to the ICM, we must account for any power contribution from inherent noise in the maps. In principle this can be done by "debiasing" the power spectrum (as will be described in Section~\ref{sec:XRPS}), but a more direct method is to "halve" the data and take a cross-spectrum \citep[e.g.~see][]{khatri2016}. However, instrumental noise can still "leak" through via such a cross-spectrum. In order to counter this, we calculate cross-spectra of noise realizations, which have amplitudes $\lesssim1/10$ the amplitudes of signal cross-spectra and, in effect, debias the cross-spectra. We perform both methods on the SZ data and present the results of the cross-spectra calculations in Figure~\ref{fig:dy_ms}.
    For the cross-spectra calculation, we take halving to be the generation of two maps covering the same area, each with half of the weight of a "full" map. 
    
    Division in half is not a trivial endeavour as these scans were taken over seven nights of observations, and even the nights with the best observing conditions had some variation in weather conditions. As such, we opt to create two halves randomly, 100 times. Cross spectra are calculated on these 100 pairs and the presented values are taken as the mean of the resultant spectra with their associated standard deviations. The 2D amplitude spectra, $A_{\text{2D}}$, for the MUSTANG-2 data are shown in Figure~\ref{fig:dy_ms} and include corrections for the MUSTANG-2 beam (PSF; the correction is shown as the dashed grey line) and MIDAS transfer function, both of which are characterized in \citet{romero2020}. 
    

    As mentioned earlier, we also calculated spectra via the debiasing route. The spectra in each ring are statistically consistent between the two calculation methods; however, Ring 2 is statistically consistent with zero as calculated via debiasing. Similarly, the spectrum in Ring 3 has negligible significance and thus we discard it from further analysis. 


\subsection{Calculation on XMM data}
\label{sec:XRPS}

    In order to calculate the power spectra for our \textit{XMM} images, we opt to debias our spectra as calculated directly on maps of fractional residuals. A noise realization can be generated as Poisson noise realizations for each pixel with its expected value given by a model of expected counts of all relevant components. To also incorporate uncertainties from the surface brightness model itself, we take 1000 models from the MCMC chains well after the burn-in. A single Poisson noise realization is generated for each of these models. The "raw" and "noise" spectra are recorded for each, as well as their difference (i.e. a "debiased" spectrum). The mean and standard deviation of these debiased spectra are used in reported expected values and associated uncertainties.
    
    \begin{figure}
        \begin{center}
        \includegraphics[width=0.45\textwidth]{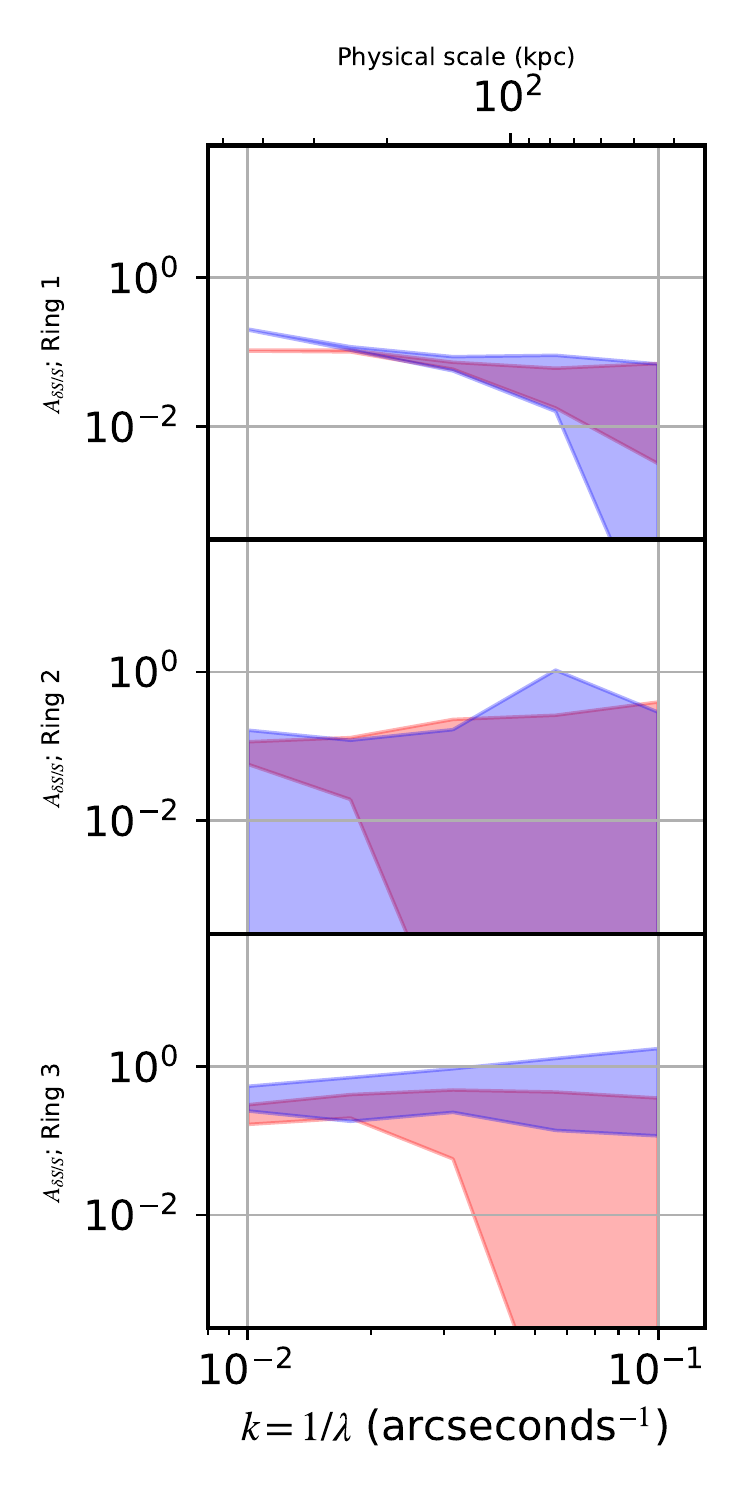}
        \end{center}
        \caption{The $\pm 1\sigma$-interval of amplitude spectra for the low energy band (red) and high energy band (blue). From top to bottom: Rings 1, 2, and 3, respectively.
        }
        \label{fig:ds_ebands}
    \end{figure}
    
    We also consider the potential contribution of faint point sources below our detection threshold. To account for these, we quantify the distribution of detected sources in our images. We normalize a LogN-LogS distribution with an index of $-1.6$ \citep{mateos2008} to our bright sources, where we take our completeness to be unity. We then randomly generate point sources of this distribution down to a minimum of 1 photon (count) when assuming a uniform (unvignetted) exposure. The final point source image, added to a noise realization, accounts for the proper (vignetted) exposure map. To stay consistent with total count expectations, we assume that the counts accumulated from these faint point sources would be equivalent to the uniform background (in count rates) in our profile fits. As such, we reduce the uniform background by the equivalent count rates.
    
\begin{figure}[!h]
        \begin{center}
        \includegraphics[width=0.45\textwidth]{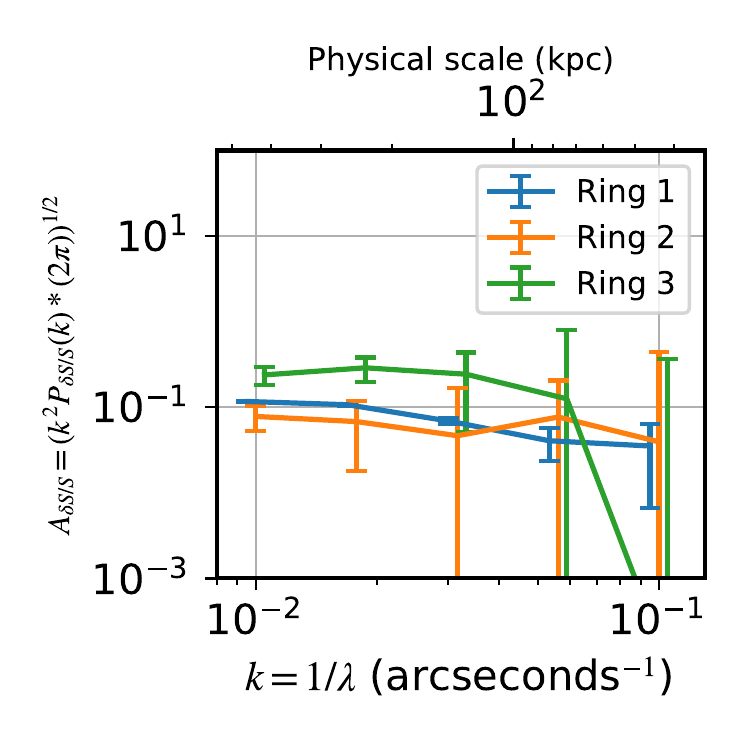}
        \end{center}
        \caption{Amplitude spectra of X-ray surface brightness fluctuations when combining both energy bands. Abscissa values are offset between rings for visual separation. Ring 2 has a similar spectrum as Ring 1 but with larger uncertainties.}
        \label{fig:ds_wtded}
    \end{figure}

    Given the general agreement between energy bands (see Figure~\ref{fig:ds_ebands}), we conclude that it is appropriate to take the weighted average of the respective power spectra, as shown in Figure~\ref{fig:ds_wtded}. When checking power spectra across individual observations and detectors, we do not find any spurious spectra. However, we also note that Figure~\ref{fig:ds_ebands} provides some insights into data quality, especially suggesting caution when attempting to interpret the combined amplitude spectrum in Ring 3 as well as the highest $k$-mode in all rings. 
    
    Both Figures~\ref{fig:ds_ebands} and \ref{fig:ds_wtded} include corrections for the PSF, which we estimate per detector, per energy band, and per ring using the \lstinline{ELLBETA} mode of the task \lstinline{psfgen}. In particular, we find the median photon energies are 800 and 2000 eV for our two energy bands, and so we estimate the PSF at those energies. For the rings, we take $x=50^{\prime\prime}, 150^{\prime\prime}$, and $250^{\prime\prime}$ and $y=0$ to be sufficient estimates of the PSFs for each ring. As in the SZ data, we see that some rings have (at least a portion of their) spectra which share the shape of the PSF correction. 
    
    To further investigate the quality in Ring 3 we calculate the radial profile (from the cluster center) of variance in the $\delta S/S$ images. We find that the average variance falls below the standard deviation of the variance (across our 1000 realizations, 3 detectors, 2 energy bands, and 3 ObsIDs) beyond $200^{\prime\prime}$. 

\subsection{3D spectra}
\label{sec:3dpsmethod}

    \begin{figure}[!h]
        \begin{center}
        \includegraphics[width=0.45\textwidth]{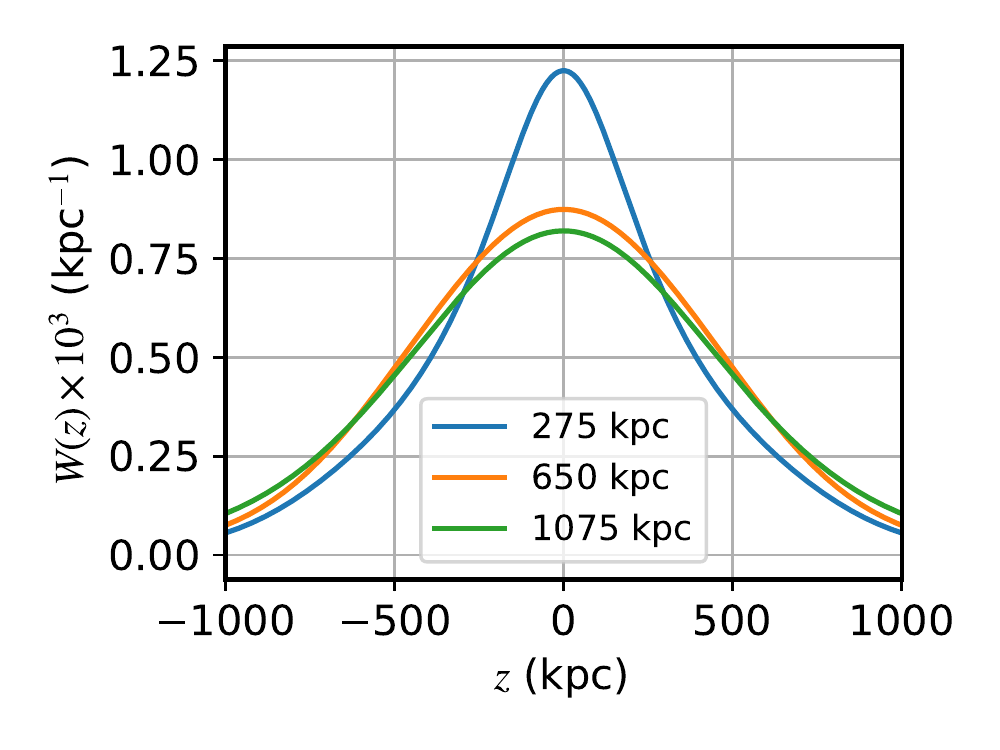}
        \includegraphics[width=0.45\textwidth]{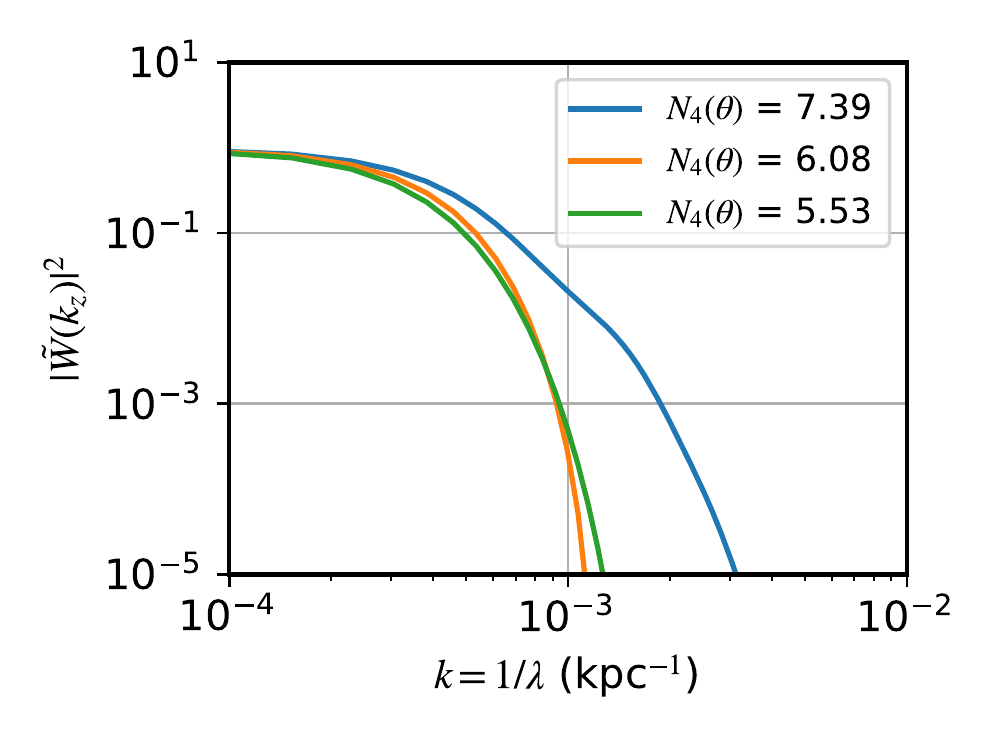}
        \end{center}
        \caption{The SZ window function in real space and in Fourier space. $N_4(\theta) = N(\theta)*1e4$ with units of inverse kpc. (see Equation~\ref{eq:window_approx}.)
        }
        \label{fig:sz_window function}
    \end{figure}
    
In this section, we relate projected 2D fluctuations to the physical 3D fluctuations by following a common  formalism \citep[e.g.][]{peacock1999,zhuravleva2012,Churazov2012,khatri2016}. 
The relation is given as:
\begin{equation}
    P_{\text{2D}}(k_\theta) = \int P_{\text{3D}}(\mathbf{k}) |\tilde{W}(k_z)|^2 dk_z,
    \label{eqn:deproj}
\end{equation}
where $z$ is the axis along the line of sight, $\theta^2 = x^2 + y^2$ is in the plane of the sky, and $|\tilde{W}(k_z)|^2$ is the 1D power spectrum of the window function, which normalizes the distribution of the relevant (unperturbed) 3D signal generation to the (unperturbed) 2D surface brightness. Additionally, $P_{\rm 2D}$ is as before, and $P_{\rm 3D}$ is the power spectrum of the 3D quantity which when integrated along the line of sight yields a surface brightness.
The SZ and X-ray window functions are respectively:
\begin{align}
    W_{\text{SZ}}(\theta,z) &\equiv \frac{\sigma_{\text{T}}}{m_{\text{e}} c^2} \frac{\bar{P}(\theta,z)}{\bar{y}(\theta)} \text{ and} \\
    W_{\text{X}}(\theta,z) &\equiv \frac{\bar{\epsilon}(\theta,z)}{\bar{S}(\theta)},
\end{align}
where $\bar{P}$ and $\bar{\epsilon}$ (emissivity), refer to the underlying 3D (spherical, unperturbed) models, which when integrated along the line of sight, produce $\bar{y}$ and $\bar{S}$, the 2D (circular, unperturbed) surface brightness models. To be sure, the relation between $\bar{S}$ and $\bar{\epsilon}$ is given by $\bar{S} = \int{\bar{\epsilon}} dz$.

Above some cutoff wavenumber, $k_{z,\text{cutoff}}$, $|\tilde{W}(k_z)|^2$ will fall off; in the regime where $k \gg k_{z,\text{cutoff}}$, we can approximate Equation~\ref{eqn:deproj} as
\begin{equation}
    P_{\text{2D}}(k_\theta) \approx P_{\text{3D}}(\mathbf{k}) \int |\tilde{W}(k_z)|^2 dk_z,
    \label{eqn:deproj_approx}
\end{equation}
where we adopt the notation used in \citet{khatri2016} and define 
\begin{equation}
    N(\theta) \equiv \int |\tilde{W}(k_z)|^2 dk_z.
    \label{eq:window_approx}
\end{equation}
In Appendix~\ref{sec:appendix_detailed_ps} we verify that this approximation in Equation~\ref{eqn:deproj_approx} is valid.
    
    \begin{figure}[!h]
        \begin{center}
        \includegraphics[width=0.45\textwidth]{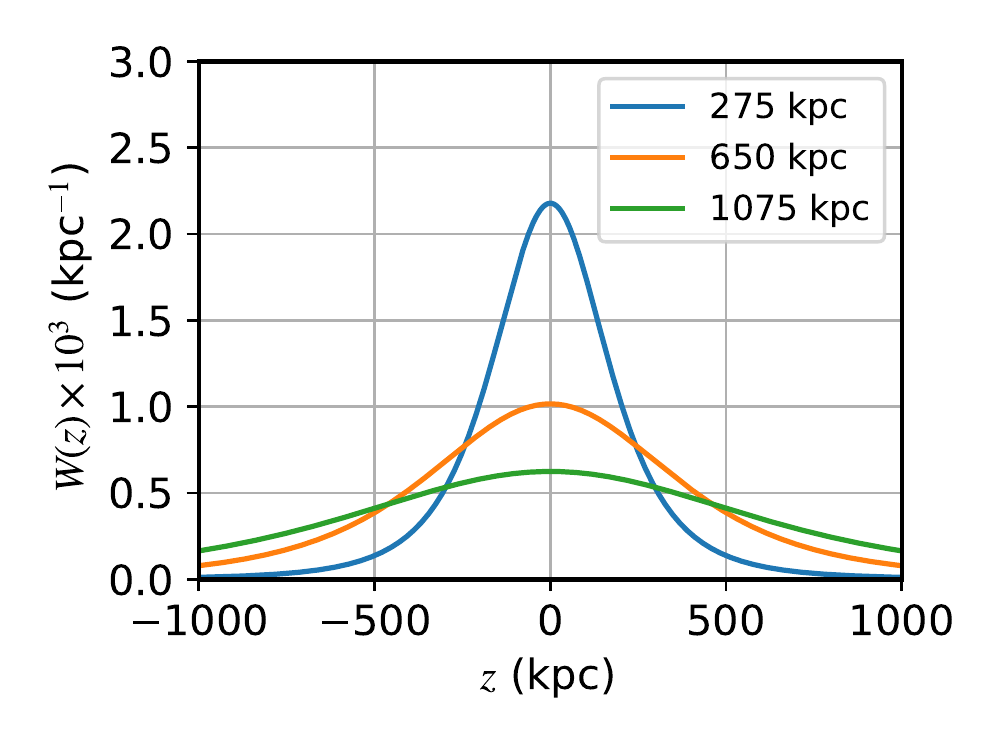}
        \includegraphics[width=0.45\textwidth]{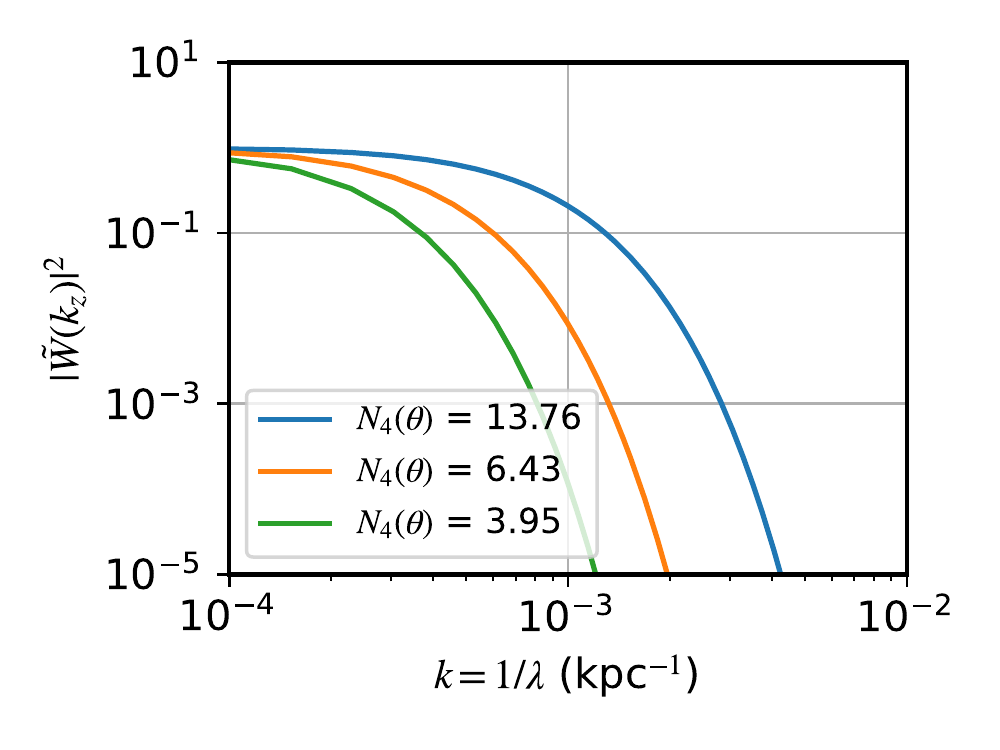}
        \end{center}
        \caption{The X-ray window function in real space and in Fourier space. $N_4(\theta) = N(\theta)*1e4$ with units of inverse kpc (see Equation~\ref{eq:window_approx}).
        }
        \label{fig:xray_window function}
    \end{figure}

The dependence of the window function on the cluster-centric radius, $\theta$, presents an issue of how to deproject over an area (e.g. over a given annulus). We therefore calculate $N(\theta)$ along many points in the range $0\arcsec \leq \theta \leq 300\arcsec$ and calculate an area-weighted average of those values (within a given annulus). Window functions (and their Fourier transform) are shown in Figures~\ref{fig:sz_window function} and ~\ref{fig:xray_window function}; the radii chosen are the effective radii for each annulus (i.e. where $N(\theta_{\text{eff}}) = \langle N(\theta) \rangle$ for $r$ in a given annulus.)

In the SZ case, this deprojection to 3D fluctuations lets us immediately arrive at pressure fluctuations ($\delta P/P$) because it is the thermal electron pressure that is being integrated along the line of sight. However, in the X-ray case, we have only derived a means of converting to fluctuations in emissivity ($\delta \epsilon / \epsilon$). Fortunately, for hot enough gas ($\sim 3$ keV), the emissivity in soft bands is weakly sensitive to temperature, and thus effectively depends only on the square of gas density, $n$. The emissivity can be expressed as $\epsilon = C n_e^2$, where we include the cooling function and mean molecular weight in $C$ and note that $C$ is weakly dependent on temperature at the temperatures of Zwicky 3146, such that $C$ acts roughly as a constant. The emissivity can be decomposed into unperturbed and perturbed terms and is linearly approximated as: $\epsilon = C n^2 [1 + 2 \delta_n]$, with $\delta_n$ being the density perturbation. This factor of 2 associated with $\delta_n$ ultimately yields a factor of 4 when relating $P_{\rm 2D}$ to $P_{3D,n}$. That is, explicitly for SZ and X-ray, we have:
\begin{align}
    P_{\delta y/y}(k_\theta) &\approx N_{\theta,SZ} P_{\delta P/P}(\mathbf{k}) \label{eqn:SZ_3D2D_approx} \\
    P_{\delta S/S}(k_\theta) &\approx 4 N_{\theta,X} P_{\delta n/n}(\mathbf{k}) \label{eqn:Xray_3D2D_approx}
\end{align}



\section{3D spectra results}
\label{sec:results}
    Given our deprojection approximation, the 3D amplitude spectra, $A_{\text{3D}}$ will simply be the 2D amplitude spectra rescaled by a scalar and multiplied by another factor of $k$. 

\begin{figure}[!h]
        \begin{center}
        \includegraphics[width=0.45\textwidth]{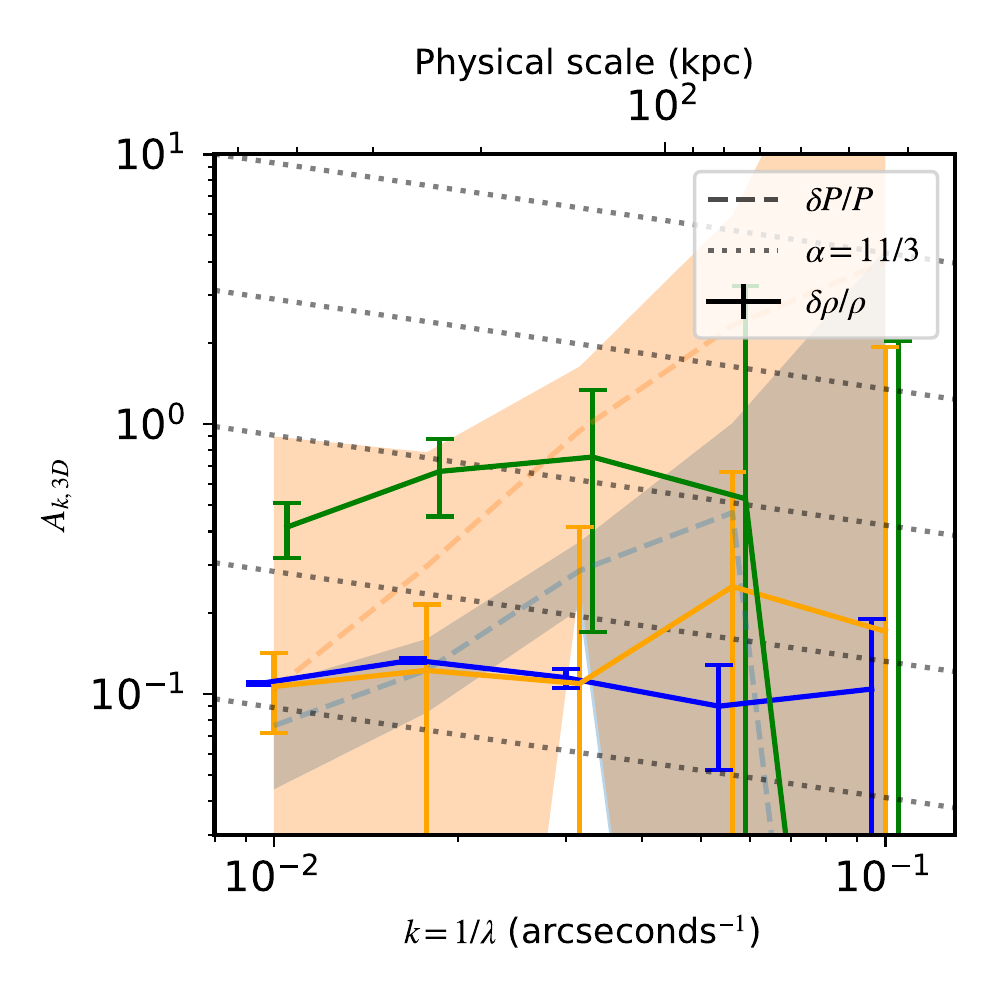}
        \end{center}
        \caption{Amplitude spectra of deprojected quantities. Colors reflect corresponding rings as in previous plots of spectra; SZ-derived spectra ($\delta P/P$) are shown as dashed lines and shaded regions while the X-ray-derived ($\delta n/n$) spectra are shown as lines with errorbars. 
        The dotted lines show the spectral indices for the power spectra (following the convention indicated in Equation~\ref{eqn:slope_convention}).
        }
        \label{fig:3dms}
    \end{figure}

As indicated in Section~\ref{sec:XRPS}, the (2D) amplitude spectrum in Ring 3 from X-ray data is likely dominated by noise. We include it in our plot of 3D amplitude spectra (Figure~\ref{fig:3dms}) and
tabulation of single spectral indices (Table~\ref{tbl:ps_products}) but do not include it in further analyses.
Similarly, we exclude Rings 2 and 3 of the SZ data from further analysis (as justified in Section~\ref{sec:SZPS}.


Figure~\ref{fig:3dms} shows the resultant density and pressure fluctuations.

If a clear peak were present in a given spectrum, we could take the amplitude at the peak ($A_{\text{3D}}(k_{\text{peak}})$) to be the amplitude of the amplitude spectrum. However, as an example, taking the highest $k$ point for Ring 2 (orange) in Figure~\ref{fig:3dms} is also problematic as it is consistent with zero. That is, choosing a peak is not solely a question of the shape of the spectra, but also of data quality. We wish to select the highest point with some threshold significance; in particular we adopt $3\sigma$ as our threshold significance. The maximum values with at least $3 \sigma$ significance are reported in Table~\ref{tbl:ps_products}. With this adopted significance threshold, we find peaks in the range $0.01 < k <0.03$, which corresponds to injection scales, $\ell_{\text{inj}}$, of $140 \text{kpc} < \ell_{\text{inj}} < 440$ kpc.

Though we may expect a changing power law (as in Figure~\ref{fig:toyspectrum}), we fit a single power law to our power spectra, omitting $k=0.01$ and report the (logarithmic) slope, $\alpha$, in Table~\ref{tbl:ps_products}, where we use the convention:
\begin{equation}
    P(k) = P_0 k^{-\alpha} \label{eqn:slope_convention},
\end{equation}
with $P_0$ being a normalization of the fitted slope. We note that without a clear indication that we are sampling below an injection scale, our slopes are not indicative of the cascade of motions to smaller scales. Moreover, with our best estimate of the injection scales ($140 \text{kpc} < \ell_{\text{inj}} < 440$ kpc), our constraints on the slope on smaller scales is minimal. These slopes do permit us to comment on the validity of our deprojection approximation (see Appendix~\ref{sec:appendix_detailed_ps}). We can additionally integrate the power spectra to obtain a measure of the variance of fluctuations; for the 3D spectra this is given as:
\begin{equation}
    \sigma_{\text{3D}}^2 = \int P(k) 4 \pi k^2 dk.\label{eqn:variance_from_PS}
\end{equation}
We report the values of $\sigma_{\text{3D}}$ in Table~\ref{tbl:ps_products}.

\begin{table}[!h]
    \centering
    \begin{tabular}{c c| c c c c c}
          & & $\alpha_k$ & $A_{\text{3D}}(k_{\text{peak}})$ & $\sigma_{\text{3D}}$ & $k_{\text{peak}} (\arcsec^{-1})$ & $\lambda_{\text{peak}}$ (kpc) \\
          \hline
        \multirow{2}{*}{Ring 1} & $\delta \rho/\rho$ & $2.5 \pm 0.1$ & $0.13 \pm 0.003$ & 0.15 & 0.02 & 250 \\
                                & $\delta P/P$    & $0.6 \pm 0.8$ & $0.29 \pm 0.08$ & 0.33 & 0.03 & 140 \\
                                \hline
         Ring 2 & $\delta \rho/\rho$ & $2.2 \pm 1.6$ & $0.11 \pm 0.03$ & 0.18 & 0.01 & 440 \\
                                \hline
       Ring 3 & $\delta \rho/\rho$ & $1.7 \pm 1.0$ & $0.67 \pm 0.21$ & 0.83 & 0.02 & 250 \\
                                \hline
    \end{tabular}
    \caption{Inferred spectral indices (logarithmic slope) and peaks of the amplitude spectra. The spectral indices assume a single power law across our sampled range with the exception of points at $k=0.1$ arcsec$^{-1}$ (we omit points at $k=0.1$ arcsec$^{-1}$). The peaks of amplitude spectra are taken with a signal-to-noise cut of 3.}
    \label{tbl:ps_products}
\end{table}

\section{Discussion}
\label{sec:discussion}
    In the context of expected amplitude spectra (see Section~\ref{sec:intro} and Figure~\ref{fig:toyspectrum}), our recovered spectra do not clearly identify an injection scale and subsequent cascade. From Figure~\ref{fig:3dms}, we may loosely infer an injection scale $100 \text{ kpc} \lesssim l_{\text{inj}} \lesssim 300 $ kpc for Rings 1 and 2. In the core an injection scale around 50 kpc could be plausible as \citet{vantyghem2021} find evidence in \textit{Chandra} data for cavities with diameters $\lesssim 50$ kpc in Zwicky 3146. Hydrodynamical simulations of AGN feedback also support such kind of injection scales (e.g., \citealt{wittor2020}).
However, the evidence for these cavities does not extend to Ring 2. In Ring 1 we see the density fluctuations increase relative to the pressure fluctuations at the larger scales probed ($\sim400$ kpc) which is consistent with a sloshing core. This also highlights that there may be multiple injection mechanisms (and scales) present in clusters.

In the present study, we refrain from making physical inferences regarding the slopes of the spectra. We do, however, compare the pressure and density spectra (in Ring 1; see Figure~\ref{fig:eos_regimes}) as well as infer Mach numbers from our spectra. We note that \citet{hofmann2016} has performed a fluctuation analysis, though not in the Fourier domain, of a sample of clusters which includes Zwicky 3146. Their analysis probed Zwicky 3146 using \textit{Chandra} data out to $r \lesssim 90\arcsec$ and can thus be compared to results from our Ring 1. They derive standard deviations for $\delta P/P$ and $\delta \rho/\rho$ of 0.004 and 0.159, respectively.\footnote{The value for $\delta P/P$ ($dP/P$ in their notation) that \citet{hofmann2016} report in their table is surprisingly low given the scatter evident in their pressure profile.} Our respective derived quantities ($\sigma_{\delta P/P}$ and $\sigma_{\delta \rho / \rho}$) are 0.33 and 0.15. Our integrated density fluctuation is in good agreement with that from \citet{hofmann2016}; however, our integrated pressure fluctuation is considerably larger than those from \citet{hofmann2016}.


\subsection{Thermodynamic state}

\begin{figure}[!h]
    \begin{center}
    \includegraphics[width=0.45\textwidth]{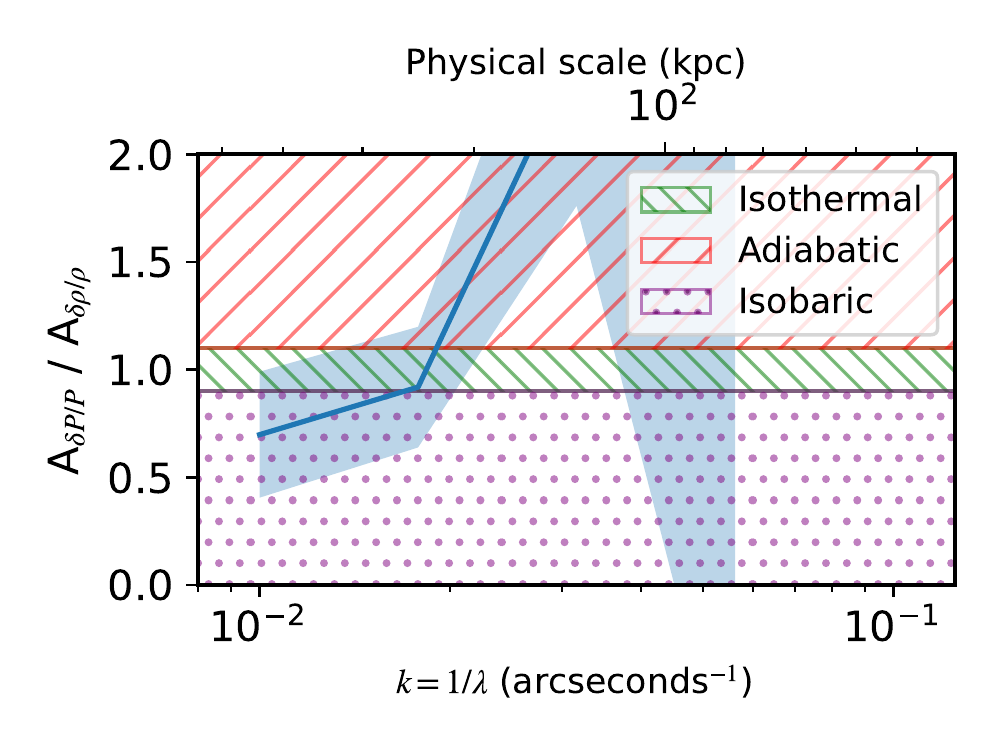}
    \end{center}
    \caption{Constraints on the thermodynamical regimes within Ring 1 given the ratio of the 3D amplitude spectra (pressure relative to density). The isothermal regime is taken to be between 0.9 and 1.1 with the adiabatic regime taken to be values above 1.1 and isobaric regime to be values below 0.9
        }
    \label{fig:eos_regimes}
\end{figure}

There are three effective thermodynamical regimes to constrain:
\begin{align*}
    \text{\textbf{adiabatic: }} \left| \frac{\delta K}{K} \right| &\sim 0, \\
    \text{\textbf{isothermal: }} \left| \frac{\delta T}{T} \right| &\sim 0, \\ 
    \text{\textbf{isobaric: }} \left| \frac{\delta P}{P} \right| &\sim 0,
\end{align*}
where $K$ is the gas entropy. 
With $\gamma$ the classic adiabatic index, we have the following relations between pressure and density in the respective regimes:
\begin{align}
    \text{\textbf{adiabatic: }} \left| \frac{\delta P}{P} \right| &= \gamma \left| \frac{\delta n}{n} \right| \label{eqn:adiabatic} \\
    \text{\textbf{isothermal: }} \left| \frac{\delta P}{P} \right| &= \left| \frac{\delta n}{n} \right| \label{eqn:isothermal} \\
    \text{\textbf{isobaric: }} \left| \frac{\delta P}{P} \right| &\ll \left| \frac{\delta n}{n} \right| \label{eqn:isobaricl}. 
\end{align}
Assuming $\gamma = 5/3$ for a monatomic gas, we can roughly divide these regimes as shown in Figure~\ref{fig:eos_regimes}. The isobaric regime $A_{\delta P/P} < A_{\delta n/n}$ is only observed at the largest scales. This is consistent with the slow perturbations driven by sloshing. Interestingly, we see that the inferred thermodynamical regime shifts to isothermal and adiabatic toward the intermediate scales. The transition from isobaric to the adiabatic state is a sign of more vigorous motions (see \citealt{Gaspari2014_PS}) as we approach the potential injection scale peak  at a few tens of kpc.
It is important to note that the isothermal transitional regime does not necessarily imply strong thermal conduction or cooling, but is a sign of a change in the effective equation of state likely due to the varying kinematics at different scales. For instance, Spitzer-like thermal conduction would substantially suppress also the density fluctuations up to hundreds kpc scale (\citealt{Gaspari2014_PS}), thus generating amplitude spectra with a very steep negative slope in logarithmic space. Our results are also in line with other observational studies \citep{arevalo2016,zhuravleva2018} which find a mixture of gas equations of state, where \citet{zhuravleva2018}, specifically analyzing a sample of cool-core clusters, find that the gas tends to be isobaric.


\subsection{Mach numbers}
\label{sec:machNumbers}

In principle we can infer non-thermal pressure, $P_{\text{NT}}$, support and ultimately a hydrostatic bias, usually defined as
\begin{equation}
    b \equiv 1 - M_{\text{HSE}}/M_{\text{tot}},
    \label{eqn:mass_bias}
\end{equation}
from our amplitude spectra presented in Section~\ref{sec:results} where we make the assumption that the non-thermal pressure support comes from (quasi) turbulent gas motions. For a perturbation with injection scale of 500 kpc, we have a simple approximation from \citet{Gaspari2013_PS} which gives us:
\begin{equation}
    \mathcal{M}_{\text{3D}} \approx 4 A_{\rho}(k_{\text{peak}}) \approx 2.4 A_{P}(k_{\text{peak}}).
\end{equation}
This can be generalized to $\mathcal{M}_{\text{3D}} \approx c_\rho A_{\rho}(k_{\text{peak}}) \approx c_P A_{P}(k_{\text{peak}}),$ where $c_\rho$ and $c_P$ have a very weak dependence on the injection scale ($\propto \ell_{\text{inj}}^{-\alpha_H}$, with $0.2 \lesssim \alpha_H \lesssim 0.3$). For an injection scale of 250 kpc, $c_\rho$ and $c_P$ will be $\sim 20$\% greater than their values for an injection scale of 500 kpc.
Other works find similar linear scalings between fluctuations and Mach numbers;
e.g., including the 3D correction $\mathcal{M}_{\text{3D}} = \sqrt{3} \mathcal{M}_{\text{1D}}$, \citet{zhuravleva2023} find a radially-averaged relation $\mathcal{M}_{\text{3D}} \approx 2.4\,\delta P/P$.

We might also consider the impact of the cool core of Zwicky 3146. Specifically, for a gas of a given Mach number we may expect density fluctuations to be significantly higher than pressure fluctuations when radiative cooling is prominent \citep[e.g.][]{mohapatra2022}. It's not clear how strong the radiative cooling is in Zwicky 3146 as the actual cooling rate may be quenched to $\sim10$\% of reported cooling flow rates \citep[see][and references therein]{romero2020}. Moreover, the cool core itself has an extent (width) of roughly 20\arcsec \citep[][]{forman2002,giacintucci2014}, so the impact of the cool core on the power spectra in Ring 1 should be negligible.

\citet{khatri2016} provide a relation between the hydrostatic bias and $\mathcal{M}_{\text{3D}}$ which we denote as $b_{\mathcal{M}}$ when derived from $\mathcal{M}_{\text{3D}}$. 

There are several limitations of our data which inhibit the goal of inferring $b_{\mathcal{M}}$ from thermodynamic fluctuations. Given the commonality of mass estimations at $R_{500}$, it is desirable to infer $b_{\mathcal{M}}(R_{500})$, but our spectra not being robust in Ring 3 does not allow us to do this. Even before then, we have the problem of estimating $\mathcal{M}_{\text{3D}}$ and eventually its (logarithmic) radial slope. As mentioned in Section~\ref{sec:results}, we cannot well determine the peaks of the spectra, both due to data quality and due to the scales accessed in this analysis. 

\begin{table}[!h]
    \begin{center}
    \begin{tabular}{c c| c c}
        \hline
          & & $\mathcal{M}_{\text{3D,peak}}$ & $\mathcal{M}_{\text{3D,int}}$ \\
          \hline
        \multirow{2}{*}{Ring 1} & $\delta \rho/\rho$ & $0.53 \pm 0.01$ & 0.32 \\
                                & $\delta P / P$    & $0.69 \pm 0.19$ & 0.80 \\
                                \hline
        Ring 2 & $\delta \rho/\rho$ & $0.43 \pm 0.14$ & 0.38 \\
                                \hline
        \end{tabular}
        \end{center}
    \caption{Inferred Mach numbers (1) based on the peak of the magnitude spectra, $\mathcal{M}_{\text{3D,peak}}$ and (2) as inferred from the integral of the spectra (i.e. variance: $\sigma^2$) and radially averaged relations in \citet{zhuravleva2023}, $\mathcal{M}_{\text{3D,int}}$.}
    \label{tbl:mach_from_ps}
\end{table}

Notwithstanding the above caveats, for spectra which we take to be robust and significant we calculate Mach numbers and report them in Table~\ref{tbl:mach_from_ps}. These values are all larger than expected for a relaxed cluster \citep[e.g.][]{zhuravleva2023}. We have deeply explored instrumental systematic errors and biases in our power spectra analyses (see Appendices~\ref{sec:powerspectracircles} and \ref{sec:appendix_detailed_ps}). We may also call into consideration the assumptions made when modelling our unperturbed cluster, e.g. would an elliptical surface brightness model be more appropriate?

\citet{khatri2016} provide a relation between the hydrostatic bias and $\mathcal{M}_{\text{3D}}$ (and attach a corresponding subscript to denote the method of calculation, $b_{\mathcal{M}}$):
\begin{equation}
    b_{\mathcal{M}} = \frac{-\gamma \mathcal{M}_{\text{3D}}^2}{3} \frac{ d \ln P_{\text{NT}}}{d \ln P_{\text{th}}} \left( 1 +  \frac{\gamma \mathcal{M}_{\text{3D}}^2}{3}\frac{ d \ln P_{\text{NT}}}{d \ln P_{\text{th}}} \right)^{-1},
    \label{eqn:mach_bias}
\end{equation}
where $\gamma$ is the adiabatic index, taken to be 5/3 for the ICM. NB that as defined in \citet{khatri2016} $b_{\mathcal{M}} \equiv M_x / M_{\text{tot}} - 1 = -b$. Following the recasting performed in \citet{khatri2016}, we find:
\begin{equation}
    \frac{ d \ln P_{\text{NT}}}{d \ln P_{\text{th}}} = \frac{ d \ln P_{\text{NT}} / d \ln r}{d \ln P_{\text{th}} / d \ln r} = 1 + 2 \frac{ d \ln \mathcal{M}_{\text{3D}} / d \ln r}{d \ln P_{\text{th}} / d \ln r}.
\end{equation}

We can employ the above equation with the average logarithmic pressure slope within Ring 1. Yet, we must also identify a logarithmic Mach number slope ($d \ln \mathcal{M}_{\text{3D}} / d \ln (r)$). Taking the weighted average of the $\mathcal{M}_{\text{3D,peak}}$ values reported in Ring 1 and the X-ray value in Ring 2, we compute a logarithmic slope. Using the weighted average of $\mathcal{M}_{\text{3D}}$ in Ring 1 we obtain $-b_{\mathcal{M}} = 0.16 \pm 0.04$. This value thus represents an estimate of the hydrostatic bias in the central region of the cluster. We note that most estimates of the hydrostatic bias are at a canonical radius like $R_{500}$, where $b$ is expected to be in the range $0.1 < b < 0.3$ \citep[e.g.][]{hurier2018}. Given the sloshing present in the core, it's plausible that the hydrostatic bias in the central region ($r < 100^{\prime\prime}$) is of similar values to values expected at $R_{500}$.

\subsection{Ellipticity}
\label{sec:ellipticity}

There is the potential for a spherical model to an ellipsoidal cluster to impart a bias on the power spectra recovered \citep[e.g.][ from perspectives of observations and simulations, respectively]{khatri2016,zhuravleva2023}. Indeed, this could apply to our result, where we should expect that our results overestimate the fluctuations at larger scales (i.e. lower $k$ modes). However, the resolution to this problem is not simple given that, much like in the Coma cluster, the ellipticities can differ between SZ and X-ray, and even between X-ray images, i.e. pn and MOS images \citep{Neumann2003}. As reported in \citet{romero2020}, the ellipticity also varies with radius. So a choice of a single ellipticity would be inherently arbitrary and would itself impart a bias at radii not matching the ellipticity chosen. By extension, employing elliptical fits to surface brightness has also been shown to sufficiently account for substructure such as a shock \citep[e.g. as in RX J1347.5-1145][]{dimascolo2019} without explicitly modeling the shock itself, hence a fluctuation analysis with such an elliptical model would risk subtracting sought-after fluctuations. Furthermore, there is no clear choice of ellipticity which escapes its own biases. Finally, when deprojecting to 3D quantities, we also introduce a degeneracy in the ellipsoidal shape and inclination of the ellipsoidal relative to the line of sight.

In a broader sense, the question can be asked: "what constitutes the unperturbed cluster model?" It should be a model that follows the shape of the gravitational potential. This question has been raised elsewhere; for example, in \citet{Zhuravleva2015} they address this by "patching" their $\beta$-model of the Perseus cluster and \citet{sanders2012} fit ellipses to surface brightness contours. In either case, this opens the question of "to what degree of complexity we should go" as well as complicating the interpretation of the underlying 3D distribution of the unperturbed thermodynamic quantities. To answer this accurately requires knowledge about the gravitational potential at a detail that is often not be available. We find ourselves in such a position: while our circular surface brightness models are likely not fully sufficient to describe the gravitational potential  we lack the data (or data of sufficient depth) to motivate another specific model other than choosing a rather arbitrary elliptical model.

\section{Conclusions}
\label{sec:conclusions}
    By leveraging our precursory multiwavelength method (\citealt{khatri2016}), in this work we have presented amplitude spectra of surface brightness fluctuations from $\delta S/S$ and $\delta y/y$ images from the X-ray (\textit{XMM-Newton}) and SZ (MUSTANG-2) data, respectively. The two instruments are well matched in angular resolution and their sensitivities are conducive to studying the intracluster medium of galaxy clusters at moderate redshift, such as Zwicky 3146 at $z = 0.29$. 

Zwicky 3146 is a relaxed, sloshing, cool core cluster. Our amplitude spectra reflect the sloshing in the core as the density fluctuations are seen to increase relative to pressure fluctuations at the largest scales in our spectra ($\sim 400$ kpc). Our amplitude spectra suggest an injection scale of $140\,\text{kpc} < \ell_{\text{inj}} < 440$ kpc. Our best constraints are in Ring 1, where the X-ray derived spectra ($\delta \rho / \rho$) suggest an injection scale of $\sim 250$ kpc, while the SZ derived spectra ($\delta P / P$) suggest an injection scale of $\sim 140$ kpc. The larger scale from X-rays reflects its sensitivity to a sloshing core. It is conceivable that the SZ data is more sensitive to fluctuations from cavities, where \citet{vantyghem2021} found potential cavities on the scale of $\sim50$ kpc; such scales are supported by AGN feedback simulations (e.g., \citealt{wittor2020}). Our comparison of pressure and density fluctuations in Ring~1 show that from large to small scales, the ICM equation of state is transitioning from isobaric to adiabatic, with a brief transition through the isothermal regime. This is another sign of increased kinematical motions (\citealt{Gaspari2014_PS}), corroborating the approach toward the turbulence injection peak potentially at a few tens kpc.

In Zwicky 3146 there is no evidence that cavities exist at moderate radii (Ring 2); and in Ring 2 we would expect an injection scale within the scales probed here. We would similarly expect an injection scale within the scales probed for our outermost ring, Ring 3. Unfortunately, neither the X-ray nor SZ data were of sufficient quality to produce reliable constraints in Ring 3. We note that in the case of SZ data an instrument with MUSTANG-2 specifications just changing the instantaneous FOV would greatly improve its ability to probe the outskirts of clusters.

Finally, we derive Mach numbers from the 3D spectra by leveraging scalings from hydrodynamical simulations. On average, we infer a turbulent 3D Mach number $\approx0.5$, with the values inferred from pressure fluctuations being relatively larger than those from density fluctuations. From the Mach numbers in the center of the cluster we infer a hydrostatic bias of $-b_{\mathcal{M}} = 0.16 \pm 0.04$.
The uncertainty in these measurements grows rapidly as one probes larger cluster-centric radii. 
Thus, future deeper and higher resolution datasets in both X-ray and SZ will be instrumental to fully unveil Zwicky 3146's kinematical state at varying radii and Fourier modes.



\section{Acknowledgements}

Charles Romero is supported by NASA ADAP grant 80NSSC19K0574 and Chandra grant G08-19117X. Craig Sarazin is supported in part by {\it Chandra} grants GO7-18122X/GO8-19106X and {\it XMM-Newton} grants NNX17AC69G/80NSSC18K0488.
MG acknowledges partial support by HST GO-15890.020/023-A, the \textit{BlackHoleWeather} program, and NASA HEC Pleiades (SMD-1726). Rishi K. acknowledges support by Max Planck Gesellschaft for Max Planck Partner Group on cosmology with MPA Garching at TIFR and Department of Atomic Energy, Government of India, under Project Identification No. RTI 4002. WF acknowledges support from the Smithsonian Institution, the Chandra High Resolution Camera Project through NASA contract NAS8-03060, and NASA Grants 80NSSC19K0116, GO1-22132X, and GO9-20109X.  LDM is supported by the ERC-StG ``ClustersXCosmo'' grant agreement 716762 and acknowledges financial contribution from the agreement ASI-INAF n.2017-14-H.0. The National Radio Astronomy Observatory is a facility of the National Science Foundation operated under cooperative agreement by Associated Universities, Inc. GBT data was taken under the project ID AGBT18A\_175. We would like to thank the anonymous reviewer for their helpful and valuable comments.

\facilities{GBT, XMM}
\software{Astropy \citep{astropy2013,astropy2018}, pyproffit \citep{eckert2017}, emcee \citep{foreman2013}, ESAS \citep{snowden2008}}

\bibliographystyle{aasjournal}
\bibliography{references}

\appendix


\section{Non-thermal pressure profile}
\label{sec:suppanalyses}

Going beyond simply calculating a hydrostatic mass bias, \citet{eckert2019} attempt to characterize the profile on non-thermal support by assuming a parameterized profile for $P_{\text{NT}}/P_{\text{tot}}$, the non-thermal pressure divided by the total pressure. One such profile proposed in \citet{nelson2014a} is given as:
\begin{equation}
    \alpha(r) = \frac{P_{\text{NT}}}{P_{\text{tot}}}(r) = 1 - A \left(1 + \text{exp}\left\{-\left[\frac{r}{B r_{200\text{m}}}\right]^{\gamma_{\text{NT}}}\right\}\right),
\end{equation}
where $A$, $B$, and $\gamma_{\text{NT}}$ are parameters fitted with respective values of $0.452 \pm 0.001$, $0.841 \pm 0.008$, and $1.628 \pm 0.019$ in \citet{nelson2014a}. With only $R_{500}$ as a node to constrain this profile, we fix $B$ and $\gamma_{\text{NT}}$ to the values from \citet{nelson2014a}. For Zwicky 3146, we obtain $R_{200\text{m}} = 2801$ kpc using the NFW \citet{navarro1996} parameters cited in \citet{klein2019}.

From \citet{eckert2019}, their assumptions yield the following relation:
\begin{equation}
    \frac{f_{\text{gas}}(r)}{f_{\text{gas,HSE}}(r)} = (1 - \alpha)\left(1 - \frac{P_{\text{th}} r^2}{(1-\alpha) \rho_{\text{gas}} G M_{\text{HSE}}} \frac{d\alpha}{dr} \right)^{-1}
    \label{eqn:ntp_relation}
\end{equation}
If we adopt the recastings $\alpha (r) = 1 - A*g(r)$ and 
\begin{equation}
    h(r) = \frac{P_{\text{th}} r^2}{\rho_{\text{gas}} G M_{\text{HSE}}},
\end{equation}
then we can arrange Equation~\ref{eqn:ntp_relation} to solve for the parameter $A$ in the non-thermal pressure fraction profile:
\begin{equation}
    A = \frac{f_{\text{gas}}(r)}{f_{\text{gas,HSE}}(r)} \left( \frac{g - h g^{\prime}}{g^2}\right),
\end{equation}
where $g^{\prime} = dg/dr$. We find $A = 0.48 \pm 0.2$.

\section{Checks and Biases with the $\Delta$-variance method}
\label{sec:powerspectracircles}
        \begin{figure*}[!h]
        \begin{center}
         \includegraphics[width=0.95\textwidth]{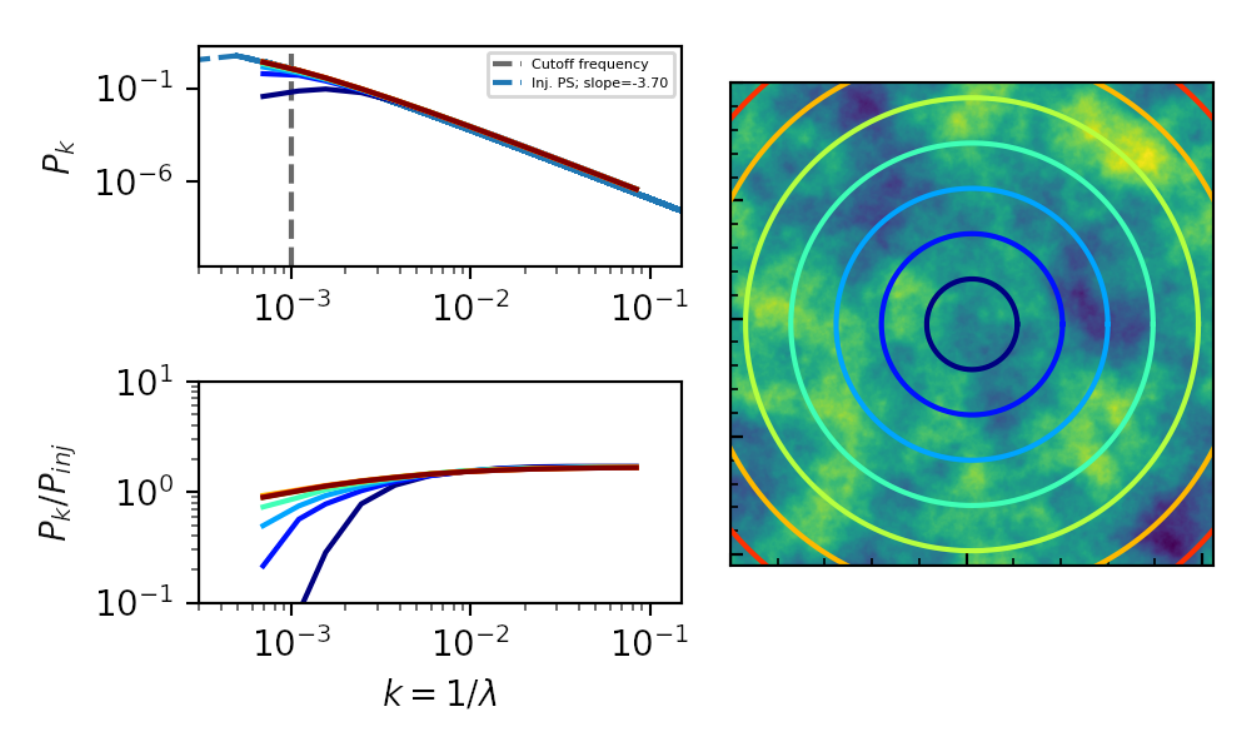}
       \end{center}
        \caption{The recovered power spectra (denoted "MH PS") of injected power spectra (blue dots, forming a thick line) with different masks ($r<$rmax) used. The color lines in the plots have corresponding circles drawn in the right panel. The MH filter quickly recovers low-k values; the cutoff wavenumber, $k_c$ is indicated by the vertical dashed line. We find very good agreement with the expected normalization bias of this method (see Appendix B in \citet{arevalo2012}). 
        }
        \label{fig:inj_noise_checks}
    \end{figure*}

Given that we wrote our own implementation of the power spectrum calculation method presented in \citet{arevalo2012}, we perform several checks to ensure we recover injected spectra as expected. In particular, we adopt an injection spectrum, $P_{\text{inj}}$ of the form:
\begin{equation}
    P_{\text{inj}} = P_0 e^{-k_c/k} k^{-\alpha},
    \label{eqn:spectral_pl}
\end{equation}
where $k_c$ is a cutoff wavenumber (towards low values), $P_0$ is a normalization, and $\alpha$ is the power law index. The value of $P_0$ is arbitrary for our checks. Similarly, noise realizations are created as images with more pixels than in our SZ or X-ray maps and the units of the pixels is arbitrary; we do check proper handling of the value of the pixel size. We test the recovery of the injected spectrum within the range of $0 \leq \alpha \leq 11/3$, where $\alpha = 11/3$ is realistically steeper than expected in 2D. We find excellent recovery of the shape (see Figure~\ref{fig:inj_noise_checks}). 

From \citet{arevalo2012}, the expected normalization bias in the recovered spectrum, $P_k$ is:
\begin{equation}
    \frac{P_k}{P_{\text{inj}}} = 2^{\alpha/2} \frac{\Gamma(n/2 + 2 - \alpha/2)}{\Gamma(n/2 + 2)},
\end{equation}
where $n$ is the number of dimensions of the data, which in our case is 2. As noted in \citet{arevalo2012}, for the 2D (and 3D) case, the bias is modest in the range $0 < \alpha < 3$, where this range encompasses expected slopes of surface brightness fluctuations. In particular, we note that the expected bias is exactly unity at $\alpha = 0$ and $\alpha = 2$. For $\alpha = 3.7$, the expected bias is 1.68 and we find a bias value of 1.66 at the highest $k$ value sampled. That is, our recovered normalization agrees very well with expectations.

\subsection{Bias for an image smoothed by a multi-Gaussian kernel}

    \begin{figure}[!h]
        \begin{center}
        \includegraphics[width=0.47\textwidth]{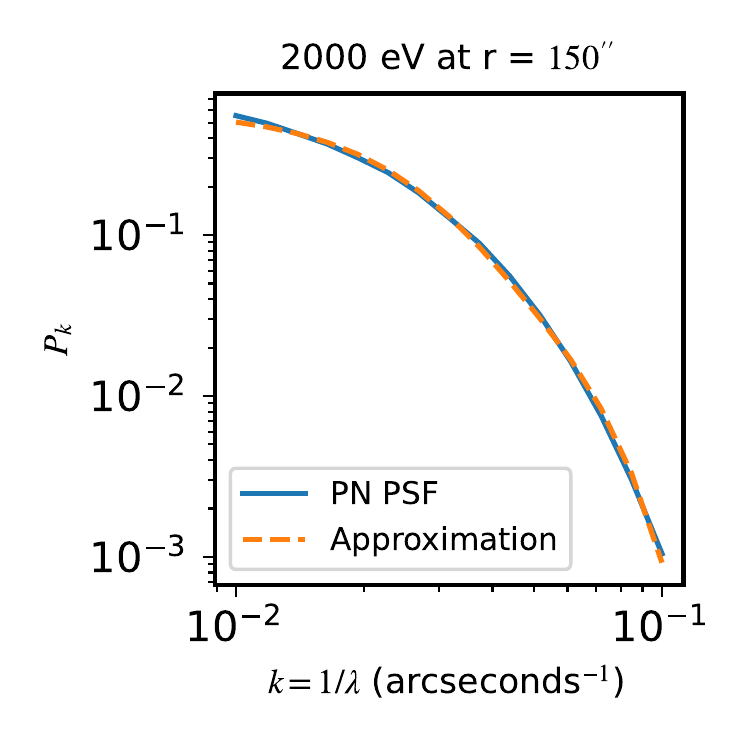}
        \end{center}
        \caption{Triple Gaussian approximation of the \textit{XMM} PSF(s). Triple Gaussians are fit to the appropriate PSF per detector, energy band, and radius; the fit is over the selected range ($0.01 < k \times {\rm arcsec} < 0.1$; we need not worry if the approximation deviates from the true PSF outside of this range.}
        \label{fig:TripleG_approx}
    \end{figure}    

    \citet{arevalo2012} derive their bias by calculating $V_{k_r}$, the variance of the filtered image as properly integrated and approximately integrated. That is, for $\Tilde{F}_{k_r}(k)$ as the Fourier transform of the filter, they evaluate
    \begin{equation}
        V_{k_r} = \int P(k) | \Tilde{F}_{k_r}(k)|^2 d^n k
        \label{eqn:variance}
    \end{equation}
    with $P(k)$ inside the integral (proper integration) and again after moving $P(k)$ outside of the integral (approximate integration) and find the ratio between the two.
    
    To derive the appropriate bias for a smoothed image, let's first note that from the convolution theorem, we have:
    \begin{equation}
        P(k) = P_{\text{u}}(k) P_{\text{PSF}}(k) = P_0 k^{-\alpha} P_{\text{PSF}}(k),
    \end{equation}
    where $P(k)$ is the power spectrum of the smoothed image, $P_{\text{u}}(k)$ is the power spectrum of the unsmoothed image, and $P_{\text{PSF}}(k)$ is the power spectrum of the PSF. In order to keep with a similar ability to integrate the expression in Equation~\ref{eqn:variance} we opt to characterize the PSF as the stack of multiple Gaussians. In our case, we'll define a radially symmetric multi-Gaussian, composed of N Gaussians, as:
    \begin{equation}
        G_{\text{multi}} = \sum_{i=1}^N c_i e^{-\mathbf{x}^2 / (2 \sigma_i^2)},
    \end{equation}
    where $c_i$ is the normalization of each Gaussian such that the total normalization is equal to unity, i.e. $\sum c_i = 1$. The Fourier transform of this multiple-Gaussian is itself a multiple-Gaussian:
    \begin{equation}
        \Tilde{G}_{\text{multi}} = \int G_{\text{multi}} e^{-i 2 \pi \mathbf{x} \mathbf{k}} d^n x =  \sum_{i=1}^N c_i e^{-k^2 / k_i^2},
    \end{equation}
    where $k_i = 1.0 / (\sqrt{2} \pi \sigma_i)$. We can further define $k_i = x_i * k_r$. If we then take $P_{\text{PSF}}(k) = \Tilde{G}_{\text{multi}}^2$, Equation~\ref{eqn:variance} now becomes:
    
    \begin{widetext}
    \begin{align}
        V_{k_r} &= \int P_{\text{u}}(k) \left[ \sum_{i=1}^N \sum_{j=1}^N c_i c_j e^{-k^2 / k_i^2} e^{-k^2 / k_j^2} \right] \left[ 2 \epsilon \left( \frac{k}{k_r} \right)^2 e^{-(k/k_r)^2} \right]^2 d^n k \\
        &=  \sum_{i=1}^N \sum_{j=1}^N 4 \epsilon^2 \int P_{\text{u}}(k) c_i c_j e^{-k^2 / (x_i k_r)^2} e^{-k^2 / (x_j k_r)^2}  \left( \frac{k}{k_r} \right)^4 e^{-2(k/k_r)^2}  d^n k.
    \end{align}
    and via the same variable recasting, we derive a new bias formulation:
    \begin{equation}
        \frac{\Tilde{P}}{P} (k_r) = 2^{\alpha/2} \left[ \sum_{i=1}^N \sum_{j=1}^N c_i c_j \left( \frac{2 x_i^2 x_j^2 + x_i^2 + x_j^2}{2  x_i^2 x_j^2} \right)^{n/2 + 2 - \alpha/2} \right] \frac{\Gamma(n/2 + 2 - \alpha/2)}{\Gamma(n/2 + 2)}.
        \label{eqn:PSFbias}
    \end{equation}
    \end{widetext}

    \subsubsection{Application to the PSFs of the EPIC cameras}

    \begin{figure}[!h]
        \begin{center}
        \includegraphics[width=0.47\textwidth]{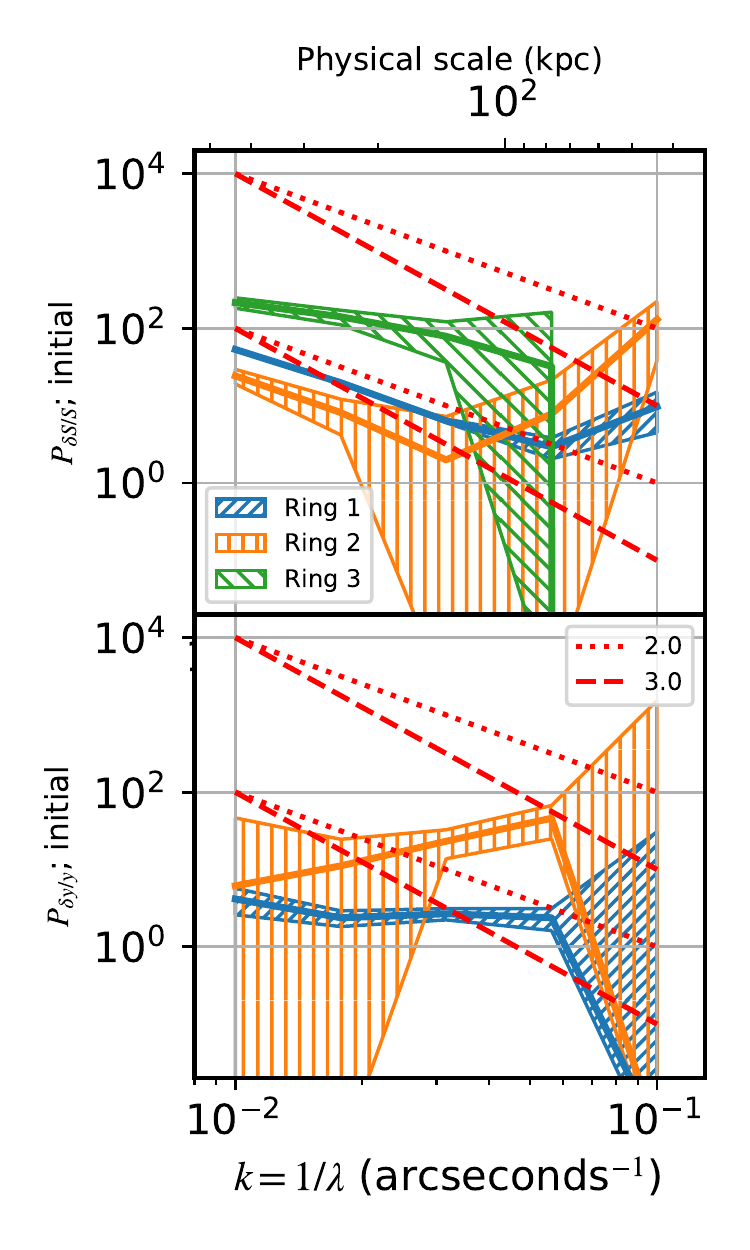}
        \end{center}
        \caption{Initial inference of $P_u(k)$ for \textit{XMM} (top panel) and MUSTANG-2 (bottom panel) without a bias correction. Dashed red lines show a power law slope $\alpha =3$  ($P \propto k^{-\alpha}$) and dotted red lines correspond to $\alpha = 2$. The steepest values we see in the X-ray spectra are $\alpha \approx 3$ and for the SZ spectra, $\alpha \approx 2$.} 
        \label{fig:InitialSpectra}
    \end{figure}
    
    In order to infer our induced biases for \textit{XMM} images, we will approximate the PSFs of our images as triple Gaussians. From Section~\ref{sec:XRPS}, we adopt a single PSF for each ring [3], each detector [3], and each energy band [2]. (Given that the three observations are well-centered on the cluster, we need not treat PSFs differently between the three observations, i.e. ObsIDs). We thus have $3 \times 3 \times 2 = 18$ different PSFs to which we fit triple Gaussians, where the fit is actually a direct fit to the power spectra of the PSFs. Figure~\ref{fig:TripleG_approx} shows that the triple Gaussian approximation stays very tight to the measured PSF; this is true for each of the 18 PSFs and respective approximations.

    For MUSTANG-2 it has been standard to calculate its beam (PSF) as a double Gaussian \citep[cf][]{romero2015a,romero2017,romero2020}.
    
    \begin{figure}[!h]        
        \begin{center}
        \includegraphics[width=0.47\textwidth]{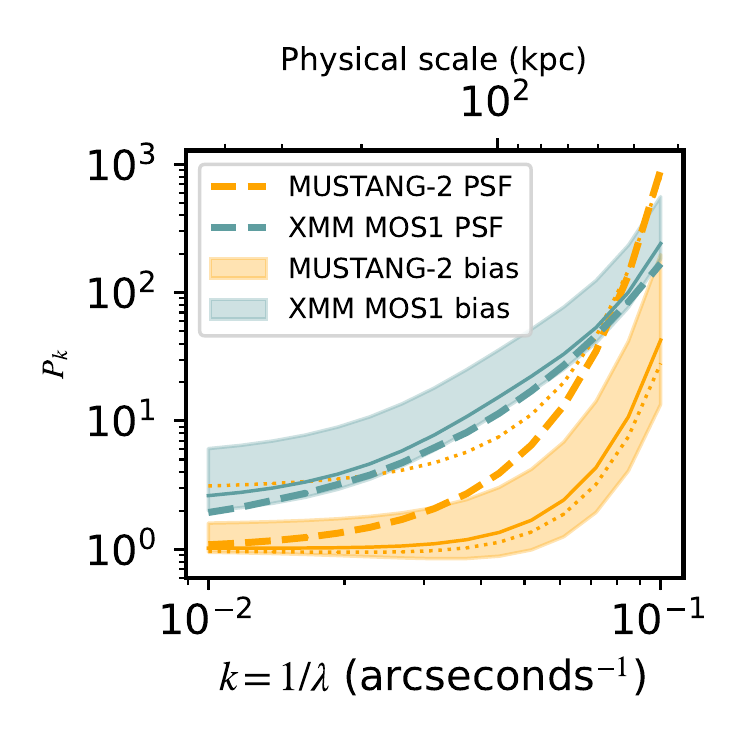}
        \end{center}
        \caption{The bias induced by PSF-convolution (shaded regions and solid lines) in comparison to the inverse of the power spectra of PSFs (dashed lines) for \textit{XMM} and MUSTANG-2. To illustrate the bias for \textit{XMM} we have only plotted the PSF of MOS1 at 800 eV bin at $r = 150^{\prime\prime}$. The solid lines indicate the bias when $\alpha = 3$ for \textit{XMM} and $\alpha = 2$ for MUSTANG-2. The shaded regions encompass the bias between $\alpha = 1.5$ and $\alpha = 4.5$ for \textit{XMM} and between $\alpha = 0.5$ and $\alpha = 3.5$ for MUSTANG-2. For a full comparison the orange dotted lines show the biases for $\alpha = 1.5$ and $\alpha = 4.5$ with the MUSTANG-2 beam.}
        \label{fig:PSFbias}
    \end{figure}
    
    We require yet another assumption before estimating our induced bias, that being the underlying spectral index. One starting point is that we could assume a Kolmogorov spectrum of $\alpha = 11/3$. However, we don't expect a single spectral index across all scales \citep[e.g.][]{Gaspari2014_PS}. At the scales probed, we should expect the slopes to only steepen towards larger $k$. Thus, we can perform an initial recovery of power spectra and identify the steepest slope between our nodes. That is, we'll expect our initial spectra to show shallower slopes towards higher $k$ due to the increasing bias.
    
    The 2D spectra from MUSTANG-2 are calculated as outlined in Section~\ref{sec:SZPS} and for \textit{XMM} as in Section~\ref{sec:XRPS}, where we note that the \textit{XMM} spectra presented in Figure~\ref{fig:InitialSpectra} are the weighted averages of the 18 different images per ring (recall there are three ObsIDs with usable EPIC data.) We find that the steepest slope in the X-ray data are steeper than $\alpha \approx 3$ while for the SZ data they are just slightly steeper than $\alpha \approx 2$. We take a slightly arbitrary uncertainty range of $\pm 1.5$ about these indices, such that for X-rays we consider biases from $\alpha = 3 \pm 1.5$ and from SZ we consider biases from $\alpha = 2 \pm 1.5$. The inferred biases are shown via solid lines and shaded regions in Figure~\ref{fig:PSFbias}.


\section{Detailed power spectra implementation}
\label{sec:appendix_detailed_ps}
    The power spectra in Section~\ref{sec:PSMethod} focus on the fractional residual maps ($\delta y / \bar{y}$ or $\delta S / \bar{S}$). However, both the power spectra from SZ (MUSTANG-2) and X-ray (\textit{XMM}) data must both have their noise bias removed. From the auto power spectrum, this would be achieved as:
\begin{align}
    P_{\delta y / \bar{y}} &= P_{\delta y / \bar{y},\text{raw}} - P_{\delta y / \bar{y},\text{noise}}, \qquad \text{or}  \\
    P_{\delta S / \bar{S}} &= P_{\delta S / \bar{S},\text{raw}} - P_{\delta S / \bar{S},\text{noise}},
\end{align}
where the raw spectra are calculated on the fractional residual, and the noise power spectrum on the associated noise realization. 

An alternative is to take a cross spectrum. To do this, we alter the standard calculation of the variance. In the framework of the delta-variance approach used, an important intermediate product is the filtered image, $\tilde{I}_{k}$:
\begin{equation}
    \tilde{I}_{k}(\theta) = \left( \frac{G_{\sigma_1} \ast I}{G_{\sigma_1} \ast M} - 
    \frac{G_{\sigma_2} \ast I}{G_{\sigma_2} \ast M} \right) M,
\end{equation}
where $M$ is the mask, $I$ is the image, and $G_\sigma$ are Gaussian kernels with corresponding widths $\sigma_1$ and $\sigma_2$. For $n$ dimensions, the variance is given as:
\begin{equation}
    V_{k} = \frac{N}{N_{M=1}} \times \int \tilde{I}_k^2 d^n x,
\end{equation}
where $N = \int d^n x$ and $N_{M=1} = \int M(x) d^n x$. In the case of calculating a cross spectrum, we have two images, $I_a$ and $I_b$, which filtered become $\tilde{I}_a$ and $\tilde{I}_b$. The variance for the cross-spectrum term is then:
\begin{equation}
    V_{k} = \frac{N}{N_{M=1}} \times \int (\tilde{I}_a \times \tilde{I}_b) d^2x,
\end{equation}
where we explicitly note that we have two dimensional images in this work.

The uncertainties in $P_{\delta y / \bar{y}}$ and $P_{\delta S / \bar{S}}$ are calculated in very similar ways. That is, if noise realizations are made to determine $P_{\delta y / \bar{y},noise}$ and $P_{\delta S / \bar{S},noise}$, where these quantities are taken as the mean power spectra across noise realizations, then the standard deviation of the respective noise power spectra can be taken as the uncertainty in the noise power spectra. Using $x$ as a stand-in for $y$ or $S$, we have:
\begin{align}
    \sigma_{P_{\delta x / \bar{x}}} & = \sigma_{P_{\delta x / \bar{x},\text{noise}}} \\
    \sigma_{A_{\delta x / \bar{x}}} & = \sigma_{P_{\delta x / \bar{x}}}/(2 A_{\delta x / \bar{x}})
\end{align}

\subsection{MUSTANG-2 error estimation}

The MUSTANG-2 map of Zwicky 3146 is comprised of 155 individual scans on source. We reprocess each scan subtracting the full model which had been fit in \citet{romero2020}. These scans span 7 observing nights.  To create 100 realizations of pairs of half maps, we randomly select half of the 155 scans to assign to "half 1" and the other half to "half 2". The cross spectra are calculated as noted above. The mean and standard deviation of power spectra are respectively taken as the expected $P_{2D}$ value and associated uncertainty.



\subsection{\textit{XMM} error estimation}

Our \text{XMM} noise realizations are fundamentally generated as Poisson noise with a model of the counts image as the mean value for each pixel. The simplest model is:
\begin{equation}
    C = \bar{S} * E + \bar{B},
\end{equation}
where $\bar{S}$ is our (smooth) ICM model (taken as a circular double $\beta$ model, see Equation~\ref{eqn:double_beta}), $E$ is the exposure map, and $\bar{B}$ is a background, which itself has multiple components. We can separate $\bar{B}$ as:
\begin{equation}
    \bar{B} = \bar{B}_p + \bar{B}_{CXB} + \bar{B}_{F} + \bar{B}_{\text{pt. srcs}},
\end{equation}
where $\bar{B}_p$ is a model of the particle background, soft proton, and for the pn detector, the OOT contribution. $\bar{B}_{CXB}$ is taken to be the uniform background (when looking at count rates) level when fitting for the ICM profile, $\bar{B}_{F}$ is the fluorescent background which has a profile proportional to the unvignetted exposure divided by the vignetted exposure, and $\bar{B}_{\text{pt. srcs}}$ is an estimate of faint point sources.

Our initial image of $B_{p}$ is formed by the addition of various image outputs from the ESAS framework and itself is subject to Poisson noise. To lessen this in the model itself, we smooth it, initially as:
\begin{equation}
    \bar{B}_p = (\frac{G_{2.5} \ast B_{\text{p}}}{G_{2.5} \ast E}) \times E,
\end{equation}
where $G_{2.5}$ is a Gaussian kernel of 2.5 pixels (pixels themselves are $2.5^{\prime\prime}$). This method should account for "losses" in chip gaps. Though this is likely sufficient, we attempt to in-fill chip gaps with the mean value of neighboring non-gap values. To do this, we iterate the smoothing; at each iteration, the non-gap values are restored to their original values, while the gap pixels are kept from the previous iteration. As the gaps are not wide, this converges quickly. Our final $\bar{B}_p$ image is obtained as:
\begin{equation}
    \bar{B}_p = (\frac{G_{2.5} \ast B_{\text{p,gap-filled}}}{G_{2.5} \ast E}) \times E.
\end{equation}

The remaining smooth background components are calculated as:
\begin{equation}
    \bar{B}_{CXB} = 10^{b_{\text{uni}}} * E,
\end{equation}
where $b_{\text{uni}}$ is the parameter fit (in logarithmic space) for the uniform background, and
\begin{equation}
    \bar{B}_{f} = 10^{b_{\text{f}}} * \frac{E_{\text{unv}}}{E} E,
\end{equation}
where $b_{\text{f}}$ is the fitted parameter (in logarithmic space) for the fluorescent background and $E_{\text{unv}}$ is the unvignetted exposure.

\begin{figure}[!h]
    \begin{center}
    \includegraphics[width=0.45\textwidth]{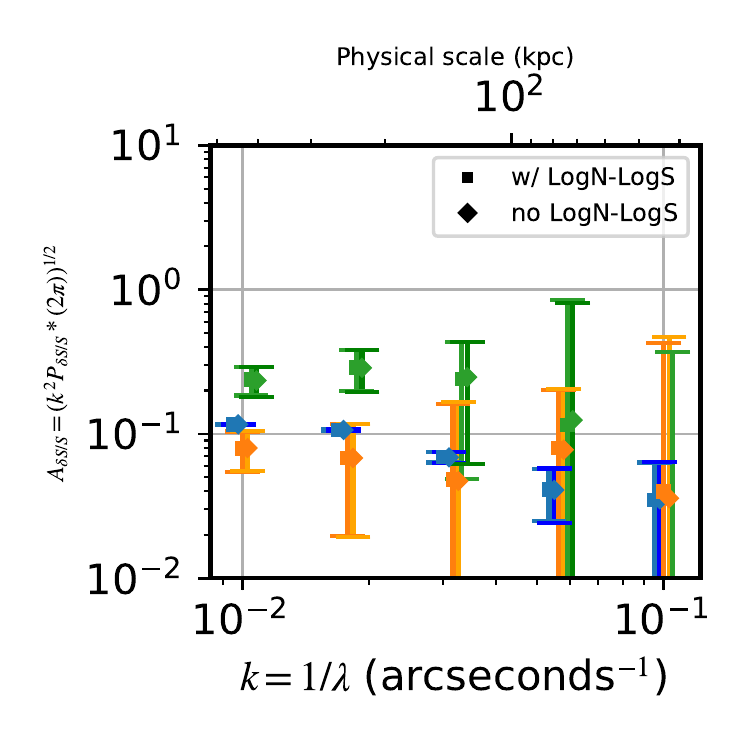}
    \end{center}
    \caption{There is a negligible difference with respect to our treatment of faint sources.}
    \label{fig:cxb_compare}
\end{figure}

If we include point sources, i.e. $\bar{B}_{\text{pt. srcs}}$, we do so as indicated in Section~\ref{sec:XRPS}. That is, we match a LogN-LogS distribution to the distribution calculated (observed) in each image using bright sources where the completeness is approximately unity and assume an index of $-1.6$. As we are taking spectra only within a circle of radius $5^{\prime}$ about the center of the cluster, we generate model point sources only in this region and assume a uniform PSF for a given detector and energy band. In particular, we adopt the PSF at $x=150^{\prime\prime}$ and $y=0$ generated by \lstinline{psfgen} as noted in Section~\ref{sec:PSMethod}. The mean photon count rate from the model point sources within $5^{\prime}$ of the cluster center is calculated and then subtracted from the same region in $\bar{B}_{CXB}$.

We wish to include covariance of our profile fit in our estimation of uncertainties in power spectra. Accordingly, we will have many models $C$, such that each noise realization is generated from an instance, $C_i$, with relevant components being dictated from chains of the profile fits. Specifically, $\bar{S}$, $\bar{B}_{CXB}$, and $\bar{B}_{f}$ depend on the chains and change with each realization.

We find that the inclusion of faint point sources barely affects our power spectra, as evidenced in Figure~\ref{fig:cxb_compare}. Even so, the results presented in this paper include an estimated contribution from such faint point sources.

\subsection{Validation of deprojection approximation}

    \begin{figure}[!h]
        \begin{center}
        \includegraphics[width=0.45\textwidth]{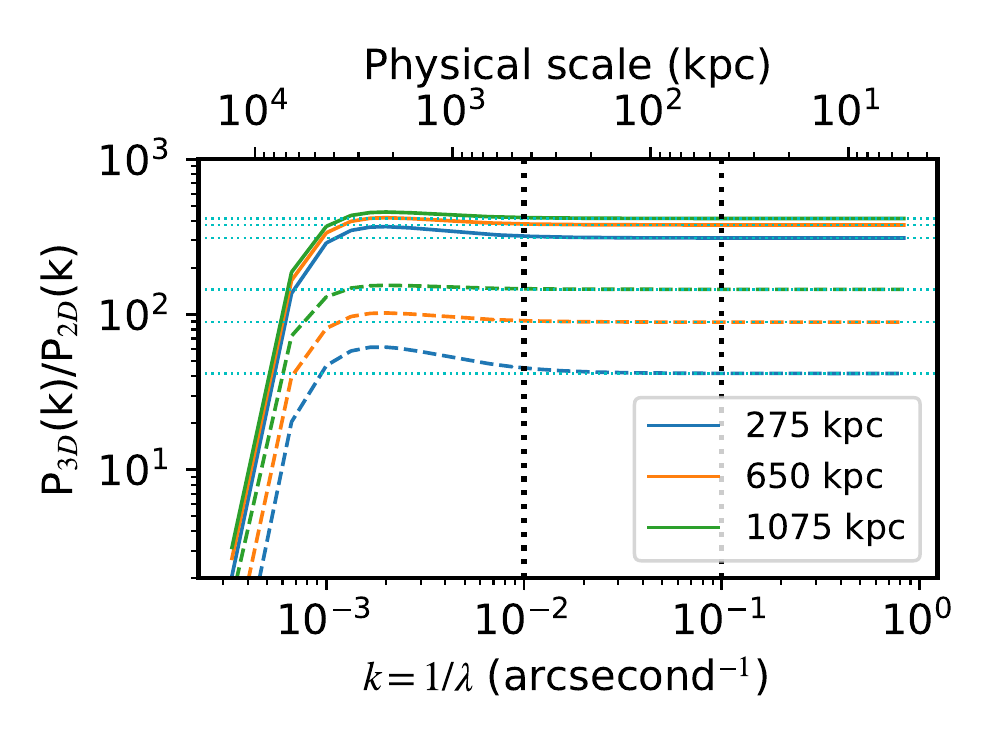}   \includegraphics[width=0.45\textwidth]{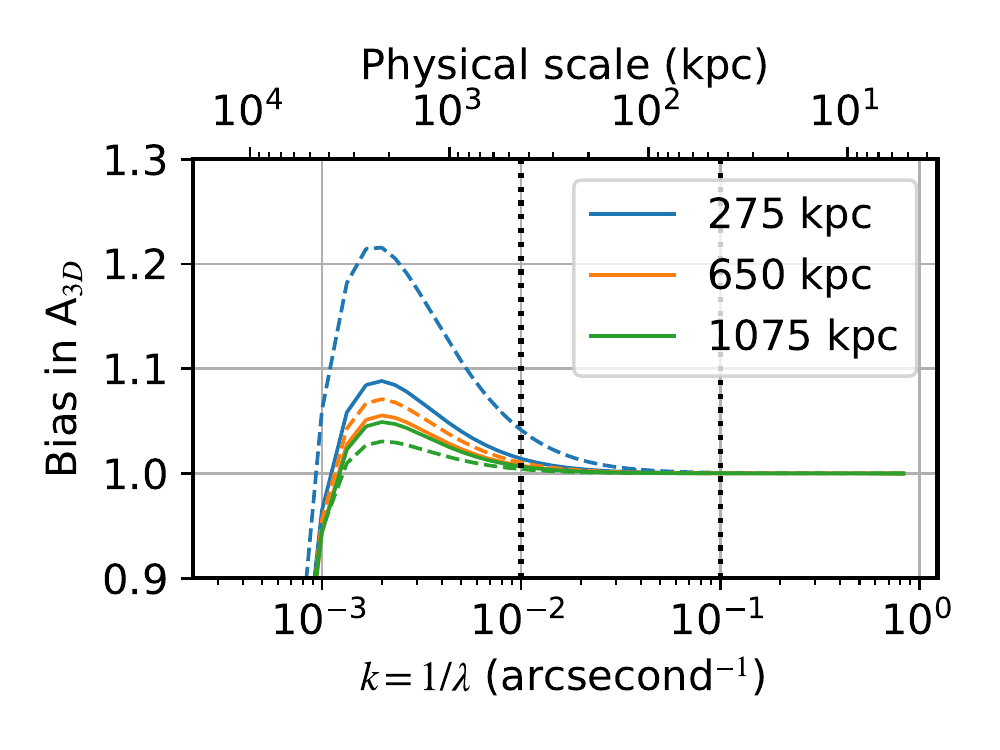}
        \end{center}
        \caption{Ratios of power spectra (above) and amplitude spectra (below) when calculated via Equation~\ref{eqn:deproj} assuming a power law distribution as in Equation~\ref{eqn:spectral_pl} with $k_c = 1e-3$ kpc$^{-1}$ and $\alpha = 3$. Solid lines show the ratios when applying the SZ window function and dashed lines show the ratios when applying the X-ray window function. The horizontal (dotted cyan) lines in the above panel show the approximate ratios between the 3D and 2D power spectra as approximated in Equations~\ref{eqn:SZ_3D2D_approx} and ~\ref{eqn:Xray_3D2D_approx}. The chosen radii are close to the effective radii (weighted averages) for $N$ of both the SZ and X-ray window functions for Rings 1, 2, and 3, respectively.
        }
        \label{fig:deproj_check}
    \end{figure}

With window functions in hand, we can easily validate our approximate power spectra deprojection (Equation~\ref{eqn:deproj_approx}) against the initial formulation, i.e. Equation~\ref{eqn:deproj}.  

The slopes found in our data (see Table ~\ref{tbl:ps_products} are generally shallower than expected, with the steepest slope being 2.4 in Ring 1. From Figure~\ref{fig:deproj_check}, we can conservatively say that we potentially underestimate the density fluctuation at $k = 0.01$ arcsecond$^{-1}$ by $\sim1.05$ (assuming the slope may be roughly 3 as we approach those scales). 

\subsection{Correlated noise on small scales}

We investigate if the increase in the amplitude spectra towards higher $k$ may be due to correlated noise on small scales. One particular investigation is the potential for secondary particles generated from collisions of cosmic rays and the telescope. If striking a detector, such secondary particles would do so effectively simultaneously. Thus, in the X-ray data, this could be seen as multiple events in a single frame. To avoid also counting multiple events from bright sources, we mask the cluster and point sources; we also perform this analysis filtering energies in our two adopted bands (400-1250 eV and 1250-5000 eV).

\begin{figure}[!h]
    \begin{center}
    \includegraphics[width=0.45\textwidth]{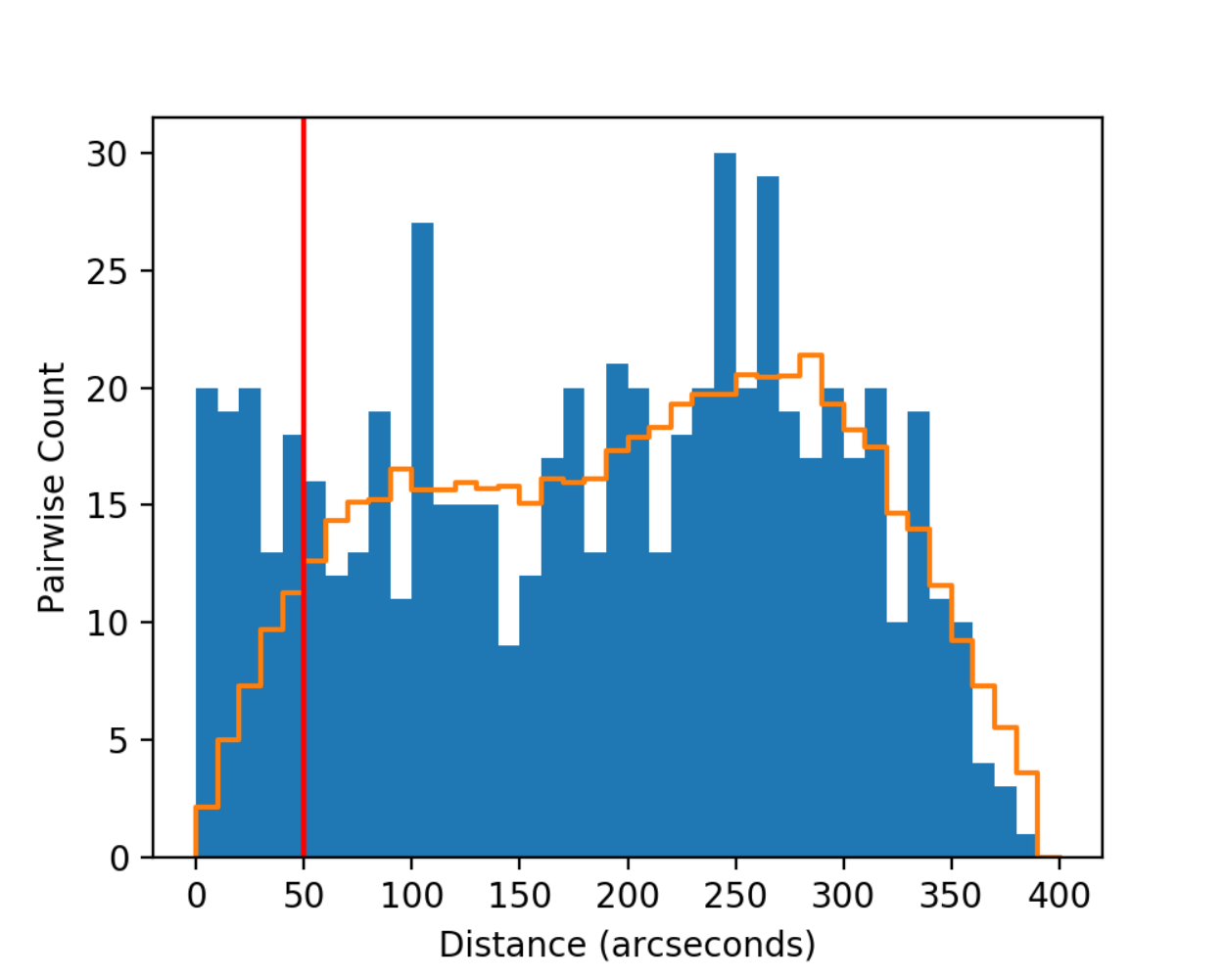}
        \end{center}
        \caption{Histogram of distances between pairs of events in the same frame as seen in our actual data (blue) and averaged over 100 realizations in which the event times are randomly shuffled.
        }
        \label{fig:pairwise_distances}
    \end{figure}
    
Wanting the shortest frame for this analysis, we analyze the pn detector from the only full frame observation (0108670101) where the time resolution is 73.4 ms. When counting the number of events per frame, we find at most three events per frame; in the high energy band only 5\% of the events in this observation occur in the same frame as another event. The distances between these events are then binned as seen in Figure~\ref{fig:pairwise_distances} (blue bars). To compare to what a random distribution would be (given our mask), we perform the same calculation but randomly shuffling the events by time (orange step plot). We consider any events occurring within $50^{\prime\prime}$ to be "short distance". The excess short distance events account for 9\% of the total pairs. We infer that the occurrence of events from (hypothesized) particle showers accounts for less than 0.4\% of high energy events in the background. Repeating this analysis for the low energy band, we find that only 0.2\% of events could be due to such particle showers. Thus, we do not find evidence that this effect could account for a rise towards higher $k$ in the recovered amplitude spectra from X-ray surface brightness fluctuations. 


\end{document}